\documentclass[bibyear]{aa}  

\usepackage{graphicx}
\usepackage{txfonts}
\usepackage{natbib,twoopt}
\bibpunct{(}{)}{;}{a}{}{,}
\usepackage[normalem]{ulem}

\usepackage[colorlinks=false, breaklinks=true]{hyperref}
\usepackage{siunitx,xspace,booktabs,twoopt,subfigure,numprint,multirow}
\usepackage{color,soul,ulem}
\usepackage{amsmath,bm}
\usepackage{orcidlink}

\newcommand{\ra}[4]{\ensuremath{#1^{\rm h}\,#2^{\rm m}\,#3\rlap{.}^{\rm s} #4}}
\newcommand{\dec}[3]{\ang[angle-symbol-over-decimal]{#1;#2;#3}}
\defcitealias{MouMacCar22}{Paper I}

\newcommand{\gtwentysixfourteen}{G026.1437-00.0420\xspace} %
\newcommand{\bdforty}{BD+43\ensuremath{^\circ}\,3654\xspace}
\newcommand{\bdfortybs}{EB27:\;BD+433654\xspace}
\newcommand{\geighteen}{G018.2660-00.2988\xspace} %
\newcommand{\bdsixty}{BD+60\ensuremath{^\circ}\,2522\xspace}
\newcommand{\bdsixtybs}{NGC\;7635\xspace}
\newcommand{\gthirtynine}{G039.2486-00.0647\xspace} %
\newcommand{\gfortynine}{G049.7065-00.1722\xspace} %
\newcommand{\gseventyfive}{G075.1730-00.5964\xspace} %
\newcommand{\hdsixteen}{EB23:\;HIP\;88652\xspace} 
\newcommand{\gonezeroseven}{G107.1371-00.6974\xspace} 
\newcommand{\gonezeroeight}{G108.0269-00.3497\xspace} 
\newcommand{\gthirtyone}{G031.6770+00.1775\xspace} %
\newcommand{\gtwentysixfifytwo}{G026.5272+00.3808\xspace} %
\newcommand{\gseventyeight}{G078.2889+00.7829\xspace} %
\newcommand{\hiptwentysix}{EB09:\;HIP\;26397\xspace} 
\newcommand{\hipthirtyone}{EB12:\;HIP\;31766\xspace} 
\newcommand{\hdfiftyseven}{EB31:\;HD\;57682\xspace} 
\newcommand{\zoph}{\ensuremath{\zeta}\;Oph\xspace} 
\newcommand{\zophbs}{EB21:\;HIP81377\xspace} 
\newcommand{\gonethreethree}{G133.1567+00.0432\xspace} %
\newcommand{\gonethreefour}{G134.3552+00.8182\xspace} %
\newcommand{\gonethreeseven}{G137.4203+01.2792\xspace} %
\newcommand{\gonefiveone}{G151.0318-00.8271\xspace} %
\newcommand{\goneonenine}{G119.4436-00.9208\xspace} %
\newcommand{\hiptwenty}{EB01:\;HIP\;2036\xspace} 
\newcommand{\sersix}{EB29:\,SER6\xspace} 

\begin{document}

  \title{A targeted radio survey of infrared-selected bow shock candidates}

  \author{
    M.~Moutzouri \inst{\ref{inst1}\fnmsep\ref{inst2}} 
    \and
    J.~Mackey~\orcidlink{0000-0002-5449-6131}\inst{\ref{inst1}\fnmsep\ref{inst2}\fnmsep\thanks{\email{jmackey@cp.dias.ie}}}
    \and
    N.~Castro~\orcidlink{0000-0003-0521-473X}\inst{\ref{inst3}}
    \and
    Y.~Gong\orcidlink{0000-0002-3866-414X}\inst{\ref{inst4}\fnmsep\ref{inst5}}
    \and
    P.~Jim\'{e}nez-Hern\'{a}ndez\orcidlink{0000-0002-6779-5181}\inst{\ref{inst6}}
    \and
    J.~A.~Toal\'{a}\orcidlink{0000-0002-5406-0813}\inst{\ref{inst7}}
    \and
    C.~Burger-Scheidlin\,\inst{\ref{inst1}\fnmsep\ref{inst2}}\orcidlink{0000-0002-7239-2248}
    \and
    M.~Rugel\inst{\ref{inst8}\fnmsep\ref{inst5}}\orcidlink{0009-0009-0025-9286}
    \and
    C.~Carrasco-Gonz\'alez\orcidlink{0000-0003-2862-5363}\inst{\ref{inst7}}
    \and
    R.~Brose\inst{9}\orcidlink{0000-0002-8312-6930}
    \and
    K.~M.~Menten\inst{\ref{inst5}\thanks{Deceased. Prof.~Karl M.~Menten passed away during the preparation of this manuscript, but was keenly involved in this project from the proposal stage and provided invaluable support throughout.}}
    }

  \institute{
   Astronomy \& Astrophysics Section, School of Cosmic Physics, Dublin Institute for Advanced Studies, DIAS Dunsink Observatory, Dublin D15 XR2R, Ireland \label{inst1}
   \and
   School of Physics, University College Dublin, Belfield, Dublin 4, Ireland \label{inst2}
   \and
   Leibniz-Institut für Astrophysik Potsdam (AIP), An der Sternwarte 16, D-14482 Potsdam, Germany  \label{inst3}
   \and
   Purple Mountain Observatory, and Key Laboratory of Radio Astronomy, Chinese Academy of Sciences, 10 Yuanhua Road, Nanjing 210023, China \label{inst4}
   \and
   Max-Planck-Institut f{\"u}r Radioastronomie, Auf dem H{\"u}gel 69, D-53121 Bonn, Germany  \label{inst5}
   \and
   Universidad Nacional Aut\'onoma de M\'exico. Instituto de Astronom\'ia. A.P. 106, 22800. Ensenada, B.C., M\'exico \label{inst6}
   \and
   Instituto de Radioastronom\'{i}a y Astrof\'{i}sica, Universidad Nacional Aut\'{o}noma de M\'{e}xico, Morelia, Mich., Mexico\label{inst7}
   \and
   National Radio Astronomy Observatory, P.O. Box 0, Socorro, NM 87801, USA\label{inst8}
   \and
   Institute of Physics and Astronomy, University of Potsdam, 14476 Potsdam-Golm, Germany
   }

  \date{Submitted 29-08-2025; Accepted 31-10-2025; This version 11-11-2025}

  \abstract
   {Bow shocks around massive stars have primarily been detected in IR emission, but radio detections are becoming more frequent with the commissioning of sensitive and large field-of-view interferometers.
   Radio data are used to probe both thermal and non-thermal emission in order to constrain the relativistic electron population and potentially particle-acceleration processes.}
   {We undertook a radio survey for bow shocks accessible from the Northern Hemisphere based on IR catalogues of candidates using the Very Large Array (VLA) and the 100 m Effelsberg Telescope. Our aim was to yield new detections and better characterise the multi-wavelength emission.}
   {We used \textit{Gaia} DR3 to re-calculate spatial motion of the driving stars with respect to the surrounding stellar population.
   We studied the radio emission from bow shocks using emission maps and spectral-index measurements and compared our results with data from catalogues and multi-wavelength emission.}
   {Of the 24 targets observed with the VLA in the 4-12\,GHz band, six were clearly detected (including two previously reported) and five were possibly detected.
   A subset of these were also observed and detected with Effelsberg at 4-8\,GHz.
   The VLA-derived spectral index maps indicate non-thermal emission for most sources, but the statistical uncertainties are large for most sources and all Effelsberg observations indicate thermal emission.
   Assuming thermal emission, we obtained upper limits on the electron density within the shocked layer. 
   We also obtained upper limits on radio emission from the bow shock of \zoph{} at a similar flux level as predictions from magnetohydrodynamic simulations.
   }
   {
   Our survey marks a significant addition to the approximately ten previously known radio-emitting bow shocks in the literature and demonstrates that deep targeted radio surveys can have a good success rate in detecting IR-selected bow shocks.
   Follow-up observations of these targets at lower and higher frequencies are encouraged to determine whether any are non-thermal emitters such as the bow shocks of \bdforty, \bdsixty{}, and LS\,2355.
   }

\titlerunning{A radio survey of bow shocks}
\authorrunning{Moutzouri et al.} 

\keywords{ 
          Stars: massive --
          Stars: winds, outflows --
          Radio continuum: ISM --
          Shock waves --
          Radiation mechanisms: non-thermal --
          circumstellar matter
          }

\maketitle

\section{Introduction}

Bow shocks arising from the hydrodynamic interaction of the interstellar medium (ISM) with stellar wind from runaway massive stars were first detected by \citet{GulSof79} using optical nebular emission lines.
The detection, however, is difficult because O and B stars are much brighter than their nebulae at optical wavelengths, and only a few optical detections can be found in the literature \citep[e.g.][]{GvaBom08, GvaMenKni14}. 

The IR wavelengths are arguably the best part of the electromagnetic spectrum to observe the bow shocks of massive runaway stars.
\citet{MeyMacLan14} showed that bow shocks are most easily detectable at mid-IR because they radiate the most energy at these wavelengths. 
This detectability can be seen clearly in the IR surveys from \citet{VanMcC88}, \citet{VanMacWoo90}, \citet{PerBenBro12, PerBenIse15}, and \citet{KobChiSch16, KobSchBal17}.
The early surveys clearly defined the different morphological and physical classes of objects and their characteristics using results from the \textit{Infrared Astronomical Satellite (IRAS)} all-sky survey.
\citet{PerBenBro12, PerBenIse15} presented results from their \textit{Extensive stellar BOw Shock Survey (E-BOSS)}, which followed up on suitable candidates from the \textit{IRAS} survey and confirmed many of their candidates to be bow shocks from massive runaway stars.
\citet{KobChiSch16, KobSchBal17} showcased over 600 bow-shock candidates at IR wavelengths, providing an excellent library of references for future observations and studies.
Using \textit{Gaia} DR3 to identify possible runaway stars, \citet{CarBenPar25} were able to detect new mid-IR bow-shock candidates with \textit{WISE} data, some of which have hints of radio emission coincident with the bow shock.

Studies at IR wavelengths are useful, but they have limitations.
In particular, since the bow shocks are excellent candidates for high-energy particle acceleration \citep[e.g.][]{DelBosMul18, DelRom12}, it is important to observe at photon wavelengths that are sensitive to both thermal emission and non-thermal emission.
This is not possible with IR because of the very strong thermal emission from dust.
At high energies (X-rays and $\gamma$-rays), even though it was thought that the non-thermal emission can be detected in the arc of the bow shocks, only upper limits have been obtained \citep[e.g.][]{SchAckBue14, ToaOskGon16, ToaOskIgn17, HESS2018_Bowshocks}.

Radio wavelengths have proven to be the best part of the electromagnetic spectrum for detecting thermal and non-thermal emission from bow shocks of massive runaway stars.
What is more, in radio we can measure gas directly and not dust.
This can be seen by the increasing numbers of detections over the past decade, including \citet{BenRomMar10},\citet{BenDelHal21}, \citet[][hereafter \citetalias{MouMacCar22}]{MouMacCar22}, \citet{VanSaiMoh22}, and \citet{VanMohCar24}.

Non-thermal emission has been detected in at least three bow shocks around the stars \bdforty \citep{BenRomMar10}, \bdsixty{} \citepalias{MouMacCar22}, and LS\,2355 \citep{VanMohCar24}.
Overall, convincing radio detections have been reported for seven bow shocks around the stars \bdforty \citep{BenRomMar10}, RCW\,49 S1 \citep{PerBenIse15, VanSaiMoh22}, NGC\,6357 G1 \citep{VanSaiMoh22}, NGC\,6357 G3 \citep{PerBenIse15, VanSaiMoh22}, \bdsixty \citepalias{MouMacCar22}, Vela X-1 \citep{VanHeyFen22}, and LS\,2355 \citep{VanMohCar24}.
A tentative detection was reported for \hdsixteen \citep{PerBenIse15, VanSaiMoh22}, and \citet{VanSaiMoh22} reported likely detections of three further bow shocks embedded in larger-scale diffuse emission.

Radio detections are primarily enabled by upgraded radio infrastructure that provide a higher sensitivity and the capacity to push the limits of observing in the radio with both wide-field and wide-band capabilities.
Recognising these new capabilities in radio astronomy, we undertook a survey of bow shocks in 2019 with the National Science Foundation's (NSF) Karl G. Jansky Very Large Array (VLA) while complementing the dataset with single-dish observations from the 100 m Effelsberg Telescope for ten of the sources in 2020 and 2021.
Following the presentation of the first results from our survey in \citetalias{MouMacCar22}, here we report on the complete survey of 24 bow shock candidates, including the sources \bdforty and \bdsixty that were discussed in that previous work.

Interest in measuring gaseous emission from bow shocks arises partly because of their potential use as laboratories for collisionless shock physics, including particle acceleration, magnetic field amplification, and non-thermal radiation \citep{DelRom12}.
They are also excellent laboratories for studying turbulent mixing at shear interfaces between the hot and warm ISM phases \citep{TanOhGro21, MacMatAli25}.
They may also be used to measure the mass-loss rates of O and B stars in cases where uncertainties from stellar spectroscopy may be very large \citep{GvaLanMac12}.
Nearby bow shocks such as that of \zoph \citep{GvaLanMac12, GreMacKav22} provide a rare opportunity to investigate hydrodynamic and radiative processes at the wind-ISM interface at high spatial resolution as constrained by observations \citep{ToaOskGon16}.
For these investigations to be successful, the gas density and ram pressure in the wind and ISM (from the reference frame of the driving star) should be well constrained by measurements.

\textit{Gaia} DR3 \citep{2023A&A...674A...1G} gives reliable distances and proper motions for a large number of Galactic runaway massive stars and, if combined with a density measurement in the bow shock, these data give the ISM ram pressure.
Computer modelling of bow shocks is advancing to the point that realistic 3D magnetohydrodynamic simulations can be run to model specific systems such as \zoph \citep{GreMacKav22} and $\lambda$ Cep \citep{SchBaaFic20} or generic studies of the properties of magnetised bow shocks \citep{MacMatAli25, DelSanPoh25}.
Similarly, synthetic observations of thermal \citep{GreMacKav22, DelSanPoh25} and non-thermal \citep{DelBosMul18, DelPoh18, DelSanPoh25} radio emission based on multi-zone or magnetohydrodynamic models are gaining predictive power.
It is therefore timely and valuable to build up a statistical sample of radio-detected bow shocks where the gas density may be measured from observations and the relative contributions of thermal and non-thermal radiation can be quantified by multi-frequency datasets.

The present work is organised as follows: Section~\ref{sec:methods} presents the selection criteria of the targets, the details of the observations with the VLA and with Effelsberg, and the methods for calculating proper motions of the driving sources of the candidate bow shocks where this was possible.
Results are presented in Section~\ref{sec:results} and discussed in the context of previous radio detections in Section~\ref{sec:discussion}.
Section~\ref{sec:conclusions} outlines our conclusions and perspectives for future follow-up observations.

\section{Methods} \label{sec:methods}

\subsection{Criteria for target selection} 
\label{sec:selection}
The main aim of our survey was to observe bow shock candidates with the VLA. Therefore we had to choose targets suitable for observing from that specific location.
We used the following catalogues of bow shock candidates:
\citet{KobChiSch16, KobSchBal17} and \citet{PerBenBro12, PerBenIse15}.
Observations were made during the 2019B Semester when the VLA was in D-configuration, choosing the C band  (4-8\,GHz) and X band (8-12\,GHz).

The biggest limitation was to choose candidates that have declination  $>-\ang{16}$ in order to minimise shadowing from the antennas\footnote{\url{https://science.nrao.edu/facilities/vla/docs/manuals/obsguide/dynsched/antenna-shadowing}} for sources that are low in the sky.
Since the D-configuration is the tightest layout for the VLA, there is a high chance the targets with declination $\le-\ang{16}$ will be overshadowed.
This limitation excluded almost half of the candidates.

Next, we had to consider the limitations on angular resolution.
The smallest resolvable angle, $\theta_\mathrm{min}$, depends highly on the observing wavelength, $\lambda$, and the arrangement of the array: $\theta_\mathrm{min} \sim \lambda/B_\mathrm{max}$, where $B_\mathrm{max}$ is the length of the largest baseline of the interferometer.
The largest observable scale, $\theta_\mathrm{max}$, is set by the beamsize of a single component telescope of the interferometer with diameter $D_\mathrm{tel}$, according to $\theta_\mathrm{max}\sim \lambda / D_\mathrm{tel}$,
For D-configuration, the observing limits (i.e,~observable angular scales of the target\footnote{\url{https://science.nrao.edu/facilities/vla/docs/manuals/oss2019B/performance/resolution}}) are between \ang{;;12} and \ang{;;240} for C band, and between \ang{;;7.2} and \ang{;;145} for the X band.
If the target is outside those limits, the array will not be able to resolve the image or will lose flux on large scales.
To decide which of the remaining objects we would include in the observation, we measured the size of the bow shocks from the available IR maps online.\footnote{\url{https://aladin.u-strasbg.fr}}

From those that had the appropriate size, we chose a sample of the brightest ones, and added three extra targets of special interest, namely the bow shocks of \bdforty, \bdsixty and \zoph.
Our final list of target bow shocks include 12 originating from O stars, 4 from B stars, and 8 from stars of currently unknown spectral type.
Of these targets, \bdforty \citep{BenRomMar10, BenDelHal21} and \hdsixteen \citep{PerBenIse15, VanSaiMoh22} have been previously detected in radio, albeit only a tentative detection for \hdsixteen.
Note that for \hdfiftyseven, a typo in Table 8 of \citet{PerBenIse15} incorrectly labelled the bow shock EB31 as being driven by the star HIP\;57862 instead of the correct star, HD\;57682 (P.~Benaglia, private communication).
The list of target bow shocks with coordinates and information on their driving stars is given in Table~\ref{tab:targets}.

\subsection{VLA observations} \label{sec:vla}
The extended capabilities of the VLA \citep{PerChaBut11}, operated by the National Radio Astronomy Observatory,  made the telescope an ideal instrument for our observations.
We investigated our sources with radio observations at \numrange{4}{12}{GHz} using the VLA (Project ID: 19B-105, PI: M.~Moutzouri), applying several techniques to analyse their spectra.
The data were reduced and analysed in the same manner as in \citetalias{MouMacCar22}, briefly summarised in the following.

The 24 targets in our sample were observed across eight separate observational runs. Each run employed a standard flux density and bandpass calibrator, along with one to four phase and amplitude calibrators. A summary of the observed targets and the associated calibrators for each run is provide in Table~\ref{tab:obsvla}.
The raw data acquired from the observations were reduced using the VLA calibration pipeline included in the Common Astronomy Software Applications \citep[CASA,][]{2007ASPC..376..127M}.
We used the CASA version 5.7 (6.1 for the pipeline) and the National Radio Astronomy Observatory's (NRAO)\footnote{The National Radio Astronomy Observatory is a facility of the National Science Foundation operated under cooperative agreement by Associated Universities, Inc.} facilities in order to reduce our data.

We imaged each target for the full frequency range of \SIrange{4}{12}{GHz}, comprising all 64 spectral windows of 128\,MHz each, using the CASA task `tclean' with the multi-scale multi-frequency deconvolution algorithm \citep{Rau11}.
The images were cleaned for a diameter of four times the FWHM of the primary beam, corresponding to a field of view of \ang{;21;} for central frequency of \qty{8}{GHz}.
We used a cell size of \ang{;;0.7} and scales sizes of [0, 5, 10, 15, 20, 25, 50, 100, 200] pixels, with a small scale bias\footnote{\url{https://casa.nrao.edu/casadocs/casa-5.5.0/global-task-list/task_tclean/about}} of 0.9.
Scales that were too large were ignored by the software.

\begin{table*}
\centering
\caption{Summary of observational runs, targets, and calibrators.}
\begin{tabular}{r >{\raggedright\arraybackslash}p{4.0cm} c >{\raggedright\arraybackslash}p{4.5cm}}
\hline
Observation ID & Observed Targets & Flux/Bandpass Calibrator & Phase/Amplitude Calibrators \\
\hline
15-10-19     & \gtwentysixfifytwo, \gtwentysixfourteen                         & 3C286 & J1804+0101 \\
16-10-19     & \hiptwentysix, \hdfiftyseven, \hipthirtyone                              & 3C147 & J0555+3948, J0632+1022, J0653-0625, J0730-1141 \\
16-10-19-2   & \bdfortybs, \gseventyfive, \gseventyeight                     & 3C48  & J2007+4029 \\
17-10-19     & \zophbs, \hdsixteen, \geighteen                         & 3C286 & J1558-1409, J1733-1304, J1822-0938, J1743-0350 \\
19-10-19     & \gonezeroeight, \gonezeroseven, \bdsixtybs               & 3C48  & J2148+6107, J2230+6946 \\
21-10-19     & \gonethreethree, \gonethreefour, \gonethreeseven, \gonefiveone & 3C48  & J0228+6721, J0349+4609 \\
22-10-19     & \gthirtyone, \gthirtynine, \gfortynine      & 3C286 & J1822-0938, J1820-0947, J1824+1044, J1922+1530 \\
23-10-19     & \goneonenine, \hiptwenty, \sersix                     & 3C48 & J0110+5632, J0019+7327, J0105+4819 \\
\hline
\end{tabular}
\label{tab:obsvla}
\end{table*}

We also used the multi-scale multi-frequency deconvolution algorithm to calculate a spectral index map by modelling the spectrum of each flux component. 
These were manually corrected for the primary beam response with the task `widebandpbcor' as in \citetalias{MouMacCar22}.
Next, we applied a mask of 5$\sigma$ (where $\sigma$ is the RMS noise level of the intensity map, quoted in Table~\ref{tab:ne}) so that we only plot the spectral index where the diffuse emission is clearly detected above the noise level.
These steps were repeated for the spectral index error maps.

\subsection{Effelsberg observations} \label{sec:effelsberg}

We conducted radio continuum imaging observations in the 4--8~GHz range towards ten selected targets using the 100 m Effelsberg radio telescope in Germany\footnote{The 100 m telescope at Effelsberg is operated by the Max-Planck-Institut f{\"u}r Radioastronomie (MPIfR) on behalf of the Max-Planck-Gesellschaft (MPG).}.
Nine of the targets were selected on the basis that we detected emission from the direction of the bow shock in a preliminary data reduction of the VLA dataset.  This includes all of our claimed and possible detections in this paper except for the sources  EB23: HIP 88652 and G075.1730-00.5964, which are marginal cases that did not show up clearly in the preliminary data reduction.  To these we added $\zeta$ Oph because its large angular extent makes it a good target for Effelsberg, with the possibility to spatially resolve the emission.
The observations were performed between 2020 September 16 and 19 (Project ID: $86-20$; PI: M.~Moutzouri). The 4.5~cm broadband secondary-focus receiver, equipped with two orthogonal linear polarisations, served as the front end, while the SPEctro-POLarimeter (SPECPOL) backend was employed. SPECPOL provides two frequency bands, covering 4--6~GHz (lower band) and 6--8~GHz (upper band), each split into 1024 channels with a channel width of 1.95~MHz.

Observations were carried out in on-the-fly mapping mode, using a scanning speed of 60\arcsec\,s$^{-1}$ and a step size of 30\arcsec\,to satisfy the Nyquist sampling criterion. Each source was mapped in both right ascension and declination directions to mitigate scanning artefacts using the basket-weaving technique \citep[e.g.][]{2017A&A...606A..41M}. The telescope focus was typically verified after sunrise and sunset, and pointing calibrations were performed every 2--3 hours using nearby bright continuum point sources. The pointing accuracy was better than 10\arcsec.

The flux density scale was calibrated using observations of 3C~286 and NGC~7027. The radio continuum data were reduced with the toolbox program and the NOD3 software package \citep{2017A&A...606A..41M}. Following the procedure described in \citetalias{MouMacCar22}, we generated continuum images at seven central frequencies that are 4.487, 4.743, 4.999, 5.383, 5.639, 6.615, and 7.639~GHz. 
Radio-frequency interference was visually inspected and manually flagged. Each image covers a bandwidth of about 200--300~MHz with typical 1$\sigma$ noise levels of $\sim$6~mJy. The angular resolution is about 140\arcsec, at 5~GHz. Of the ten sources observed, BD+43$^{\circ}$3654 and NGC~7635 were previously presented in \citetalias{MouMacCar22}, while the remaining sources are reported in this work.

\subsection{Calculation of peculiar proper motion for stars} \label{sec:motion}

We estimated the local proper motions of the targets using surrounding stars as reference benchmarks to identify any peculiar motion relative to the distribution of the nearby objects.
These calculations are based on the proper motions and parallaxes provided by \textit{Gaia} DR3 \citep{2023A&A...674A...1G}.
However, some of the faintest stars do not have \textit{Gaia} counterparts in the current release.
Additionally, the parallaxes and proper motions of certain objects exhibit large uncertainties.
These targets have values `n/a' in Table~\ref{tab:targets} for their proper motion.

The local reference system was established using all stars provided by \textit{Gaia} within 10\arcmin\ of each target.
From the projected stars within this 10\arcmin\ aperture, we selected those located within a local volume of 250 pc around the target.
Parallax was used as the distance estimate for both the target and the local reference volume.  
The local proper motions of each target, listed in Table~\ref{tab:targets} (column Peculiar PM), are with respect to the average proper motions of the selected local volume, also listed in Table~\ref{tab:targets} (column Env.~PM).

\subsection{Estimating the gas density from radio emission and upper limits} \label{sec:electrondensity}

Considering thermal bremsstrahlung emission, we can obtain a constraint on the emission measure of the bow shock and thereby on the electron density, $n_e$, by assuming smoothness and a simple geometry for the bow shock.
If we assume all of the detected emission is bremsstrahlung then we obtain an upper limit on $n_e$ because part of the emission could be non-thermal.
Non-detections are upper limits by default.

Furthermore, hot stars are surrounded by H~\textsc{ii} regions of photoionised gas, which also emit bremsstrahlung that contributes to the line-of-sight emission measure, implying that only part of the detected emission is from the bow shock.
The degree of contamination depends on the relative size of the bow shock and H~\textsc{ii} region, as well as the compression factor of the forward shock (because the recombination radiation scaled with $n^2$).

Following \citet{VanHeyFen22} and \citet{Spi78}, the bremsstrahlung emissivity, $j_\nu$, can be expressed as a function of the gas temperature, $T$; electron and ion number densities, $n_e$ and $n_i$; ionic charge, $Z_i$; and the observation frequency, $\nu$:
\begin{equation}
\frac{j_\nu}{n_e n_i} = 5.44\times10^{-39} \frac{g_{\text{ff}} Z_i}{\sqrt{T}} \exp{\left(-\frac{h\nu}{kT}\right)} \;\;\mathrm{erg\,cm^3\,s^{-1}\,sr^{-1}\,Hz^{-1}} \;,
\end{equation}
where $g_{\text{ff}}$ is the Gaunt factor, which depends on $T$, $Z_i$, and $\nu$ in the radio regime, and where $h\nu/kT\ll1$.
For the case considered below, $g_{\text{ff}}\approx5$.

We also assumed the following:
\begin{enumerate}
    \item The bow-shock emission is spatially resolved.
    \item The density within the layer of the shocked ISM is constant.
    \item Within the same shocked layer, there is a constant temperature of $T=8000$\,K, which is appropriate for isothermal shocks in photoionised gas.
    \item The bremsstrahlung emission is dominated by H$^+$ ions and electrons so that $Z_i=1$ and $n_e\approx n_i$.
    \item The path length of a ray through the shocked layer near the apex is approximately the standoff distance of the bow shock from the star, $R_0$.
    \item The emission is optically thin.
\end{enumerate}
All of these assumptions are approximately valid for the VLA observations of bow shocks, for stars moving within the diffuse ISM such that the forward shock is radiative but not so dense that radio emission becomes optically thick.
The assumptions allowed us to invert the radiative transfer equation for the specific intensity, $I_\nu = \int j_\nu d\ell$, (where $\ell$ is the path of the ray, and absorption is ignored) to obtain a crude estimate of the mean value of $n_e$ within the shocked layer, or an upper limit in case the bow shock is not detected in radio.
We obtained
\begin{equation}
    n_e = 1.80\times10^{-21} \sqrt{\frac{n_e n_i}{j_\nu(T,Z,\nu)} \frac{\mathrm{pc}}{R_0} \frac{sr}{\theta_\mathrm{B}} \frac{I_\nu}{\mathrm{Jy\,beam^{-1}}} } \;\;\mathrm{cm^{-3}} \;, \label{eq:ne}
\end{equation}
where $\theta_\mathrm{B}$ is the solid angle of the telescope beam, which is in the range $\theta_\mathrm{B} \in [1,3.5]\times10^{-9}$\,sr for our VLA observations.

For the stars with a known spectral type (17 of the 24 targets), all of which also have a measured distance, we could also estimate the pre-shock gas density from the measured standoff distance, $R_0$, given by
\begin{equation}
R_0 = \sqrt{\frac{\dot{M}v_\infty}{4\pi \rho_0 (v_\star^2 + c_s^2)}} \;,
\label{eqn:standoff}
\end{equation}
where $v_\infty$ is the terminal velocity of the stellar wind, $\rho_0$ is the ISM gas density, and $c_s\approx12\,\mathrm{km\,s}^{-1}$ is the ISM sound speed in photoionised gas.
The ISM electron density, $n_{e,0}$ can be obtained from $\rho_0$ assuming 1.1 electrons per H nucleus (if He is singly ionised) and the ISM mass fraction of H of $X_H\approx0.7$.
We have measured $R_0$ and $v_\star$ for these 17 targets using \textit{Gaia} DR3 distances and proper motions, so an estimation of $\dot{M}v_\infty$ will give $n_{e,0}$.

We estimated the stellar luminosity, mass, temperature, and radius using the tables of \citet{MarSchHil05} for the O stars, and from these we obtained the mass-loss rates by using the prescription of \citet{VinDeKLam01} (assuming solar metallicity).
Wind velocities were obtained from the escape velocity following \citet{EldGenDai06}.
For the B stars, we used the tables of \citet{Krt14} to get mass-loss rates and wind velocities for main-sequence stars and those of \citet{KrtKubKrt21} to get the same quantities for supergiants.
\gonezeroeight{} has no luminosity class, so we assumed a supergiant, and for both it and \gonezeroseven{}, we took $\dot{M}$ and $v_\infty$ for the $25\,\mathrm{M}_\odot$ calculations.
For $\zeta$ Oph, \hdsixteen, \bdforty, and \bdsixty we took mass-loss rates from literature estimates \citep{GvaLanMac12, GvaBom08, MouMacCar22, GreMacHaw19}.
The mass-loss rates and wind velocities we obtained are listed in Table~\ref{tab:app:mdot}.

\section{Results} \label{sec:results}

The 24 targeted bow shocks and bow-shock candidates are listed in Table~\ref{tab:targets}, sorted by increasing R.A.~of the driving star.
They are identified according to their name in the E-BOSS II \citep{PerBenBro12, PerBenIse15} and \citet{KobChiSch16} catalogues in the first column, and the ID of the driving star is listed in the second column.
Of the 24 targets, six were convincingly detected as extended radio sources with the VLA, five show some hints of emission, and 13 are undetected.

\begin{figure*}
    \centering
    \includegraphics[width=0.9\textwidth,trim={0cm 1cm 0cm 0cm}, clip]{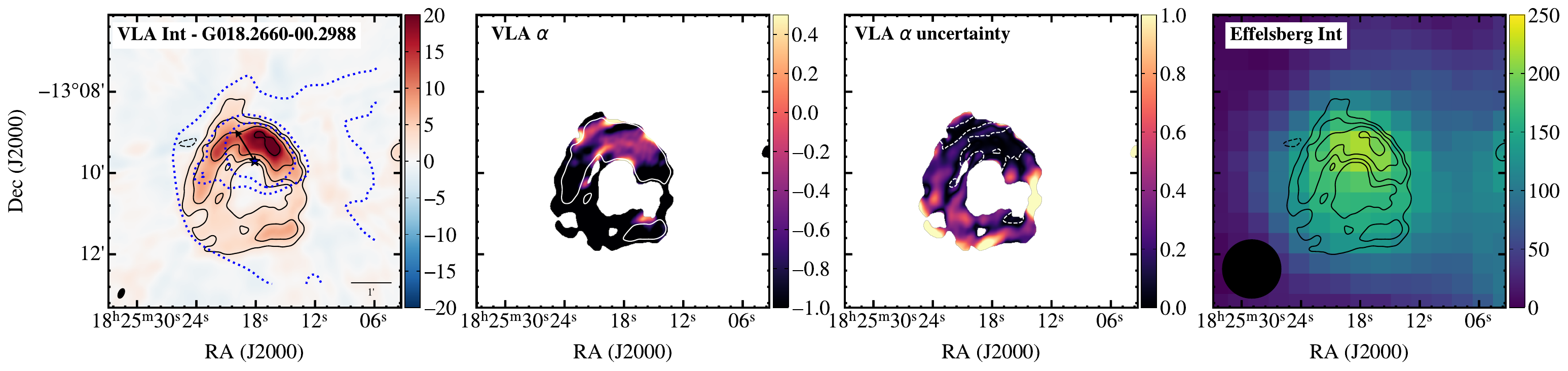}
    \includegraphics[width=0.9\textwidth,trim={0cm 1cm 0cm 0cm}, clip]{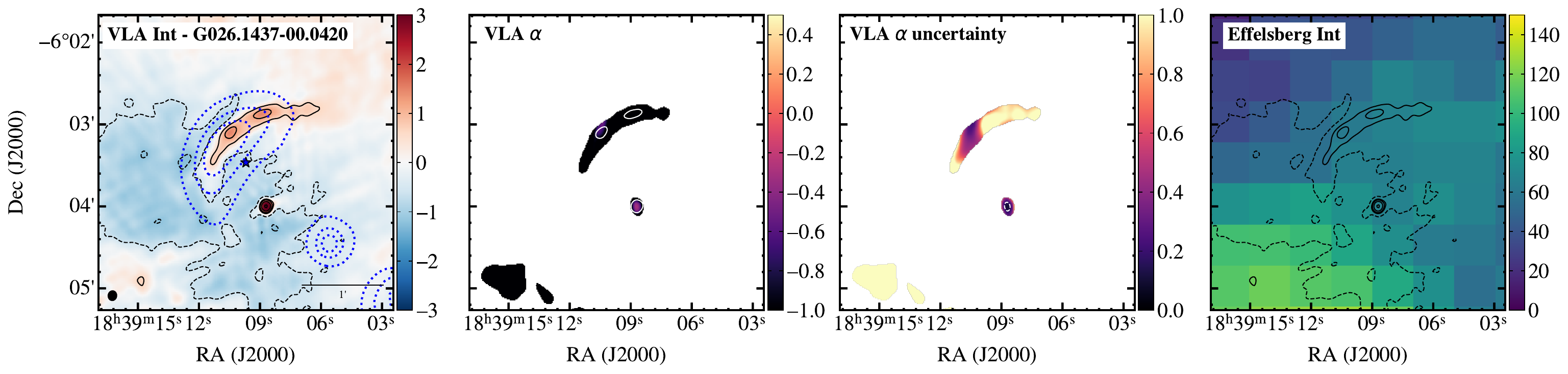}
    \includegraphics[width=0.9\textwidth,trim={0cm 1cm 0cm 0cm}, clip]{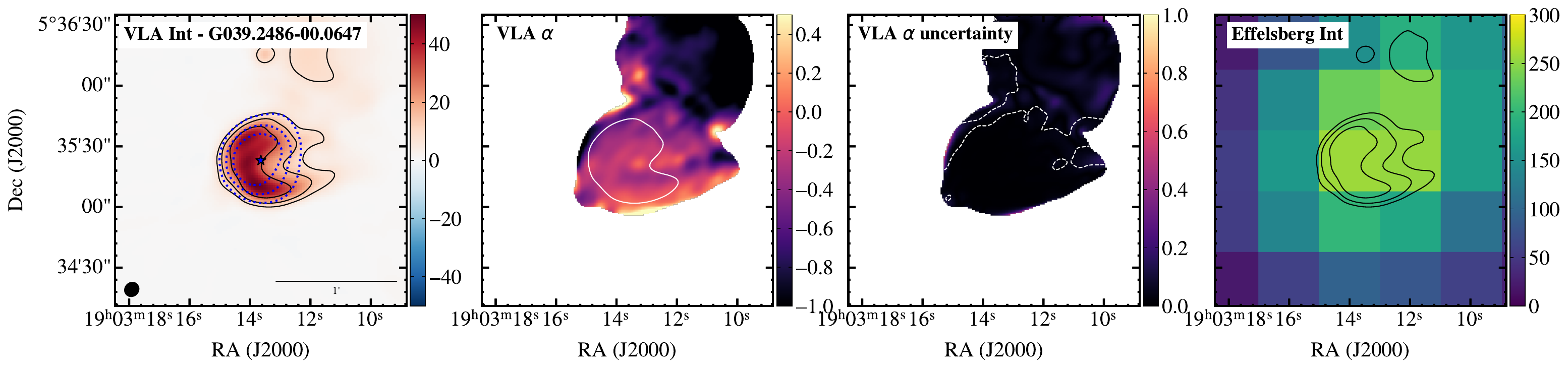}
    \includegraphics[width=0.9\textwidth,trim={0cm 1cm 0cm 0cm}, clip]{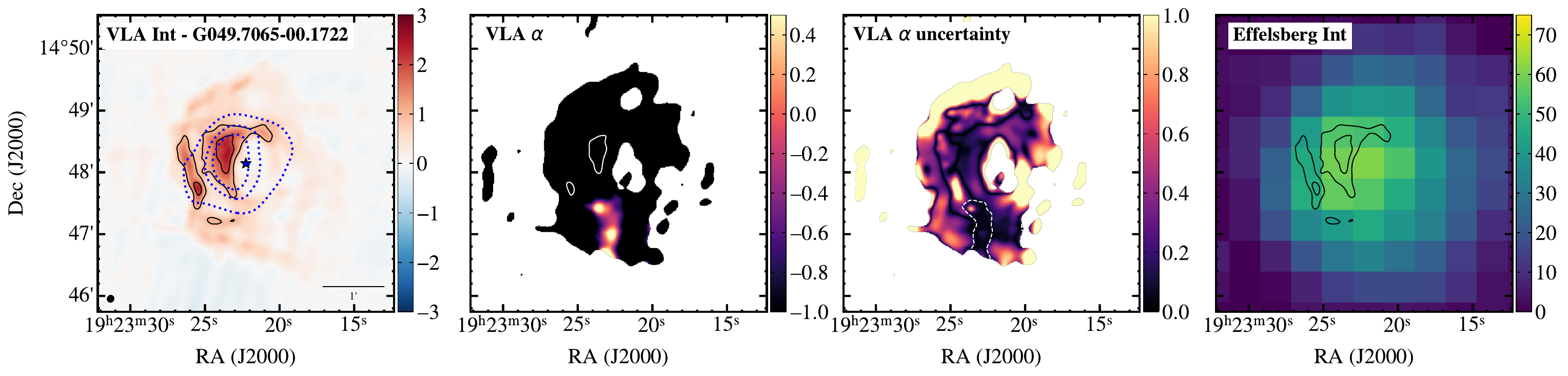}
    \includegraphics[width=0.9\textwidth,trim={0cm 1cm 0cm 0cm}, clip]{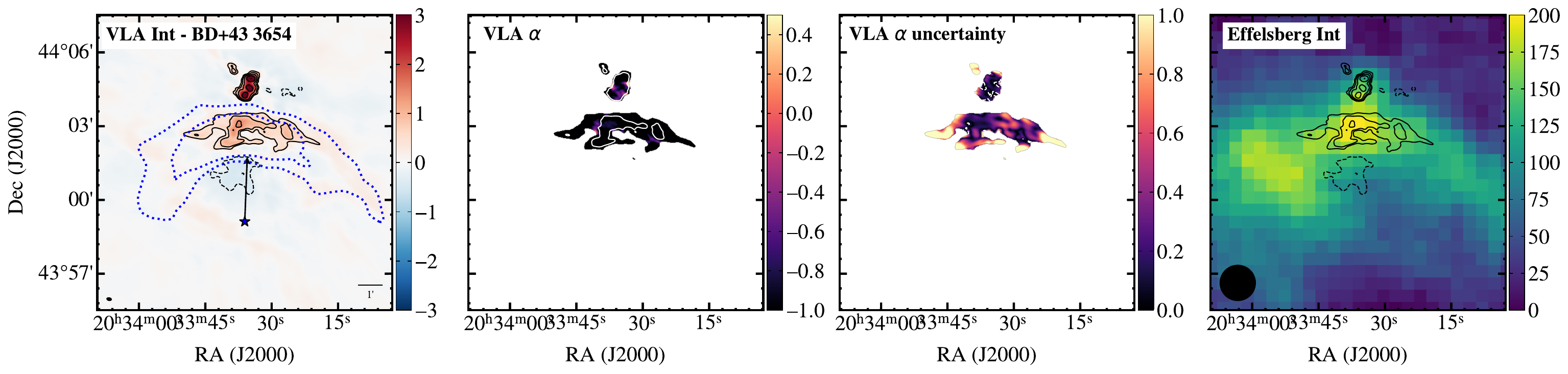}
    \includegraphics[width=0.9\textwidth]{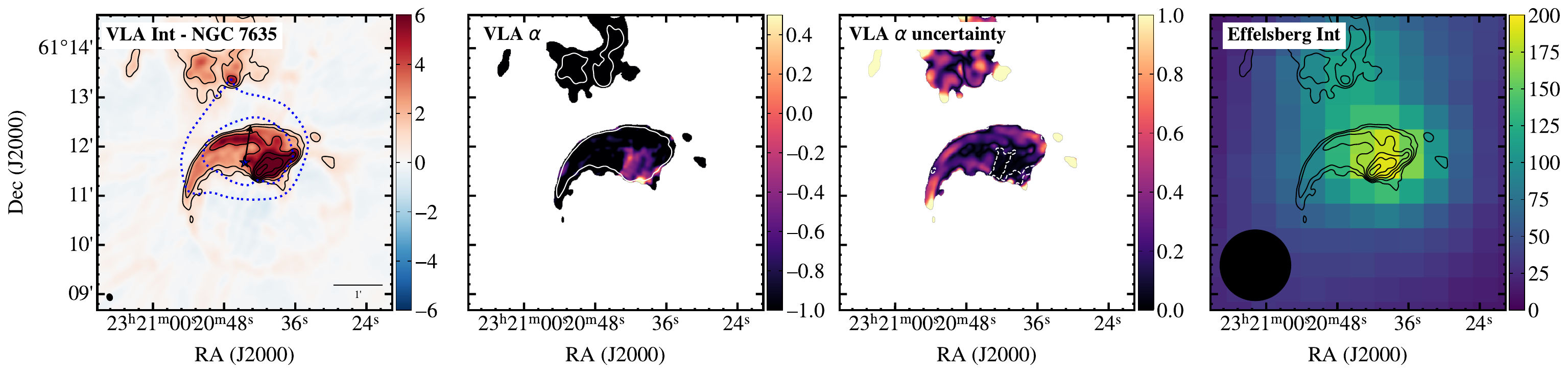}
    \caption{Presentation of the detected targets listed in Table~\ref{tab:targets}.
    Each row shows a single target, labelled in the title of the first column.
    From left to right, the panels show (1) the VLA intensity map in mJy\,beam$^{-1}$ with black contours at $[-10,-5,5,10,20,40,80,160]\,\sigma$; (2) the map of the power-law spectral index, $\alpha$, with $10\sigma$ intensity contour overlaid; (3) the map of the 1-$\sigma$ absolute uncertainty in $\alpha$, with a contour showing $\alpha=-0.5$; and (4) the Effelsberg intensity map at 7.639\,GHz (colour scale in mJy\,beam$^{-1}$) with VLA intensity contours overlaid and the beamsize shown for cases where it is not too large (in all cases, about 3 pixels in diameter).
    Also shown in the first panel are a line showing \ang{;1;} scale at the bottom right, the synthetic beam FWHM at bottom left, blue dotted contours of  WISE \SI{22}{\um} emission, and a black arrow showing the direction of peculiar proper motion, where reliably estimated.
    }
    \label{fig:detections2}
\end{figure*}

\subsection{Detections}

Intensity images from the VLA for the six clear detections are shown in Figure~\ref{fig:detections2}.
Given the small number of radio-detected bow shocks currently known in the literature, this is an excellent return of positive detections for the survey.
They are discussed one by one in the following.

\subsubsection{\geighteen}
This bow shock candidate is entry 72 in the catalogue of \citet{KobChiSch16}, driven by the O6V-O5V \citep{JiZhoEsi12} star 2MASS\,J18251808-1309427.
This star is part of a young star cluster creating the interstellar bubble N\,22 \citep{JiZhoEsi12} and the larger H~\textsc{ii} region RCW\,166.
The bow shock is observed with a width $\sim\ang{;;50}$ (shown in the third row of Figure~\ref{fig:detections2}), although it is quite possible that its features are obscured by the more extended bubble that surrounds it.
Indeed, the radio emission that we see in Figure~\ref{fig:detections2} seems to correlate better with the nebular emission in the DSS Red channel than with the IR arc, implying that we are detecting diffuse emission from the H~\textsc{ii} region as well as the putative bow shock.
The peculiar proper motion of $\approx25$\,km\,s$^{-1}$ to the NE is approximately what one would expect based on the orientation of the bow shock, but the uncertainties in the direction are quite large.

The spectral index map shows that the source emits a mixture of thermal emission and non-thermal emission (third row, second panel of Figure~\ref{fig:detections2}) although the uncertainties are again rather large (third row, third panel of Figure~\ref{fig:detections2}), particularly in regions with nominally very negative spectral index.
At the position of the IR arc, the spectral index could be consistent with both thermal and non-thermal emission, with $-0.5\lesssim\alpha\lesssim0$.
Further investigation on the north part of the source could be made, although the IR arc itself is so small and embedded in the bubble that a definitive conclusion on its nature may not be easy to obtain.
The Effelsberg image (third row, fourth panel of Figure~\ref{fig:detections2}) shows that the bubble is slightly resolved spatially, with the brighter north side clearly also brighter with Effelsberg.
Unfortunately the bow shock is much too small to be seen in the Effelsberg image and is superimposed on the H~\textsc{ii} region emission.

With the VLA we detect a clear lack of emission from the central part of the H~\textsc{ii} region, where most of the O stars in the young star cluster are located \citep{JiZhoEsi12}.
The most likely explanation is that the O stars have evacuated a low-density cavity through the combined actions of their stellar winds.
In that case the IR arc could represent the edge of the wind bubble \citep{MacHawGva16} and may not be associated with a significant gas overdensity.
Alternatively we may be seeing a young O star that has just been ejected from its birth cluster and is driving a bow shock at the edge of the cluster wind-cavity.

This young star cluster is located in a region of bright gamma-ray emission associated with pulsars and pulsar-wind nebulae \citep{HESS2020}.
Our possible detection of non-thermal emission indicates that this young star cluster could be efficiently accelerating particles \citep[cf.][]{PerCasGab24}, although high-energy radiation will be difficult to detect from this cluster on account of source confusion from the bright nearby sources.

\subsubsection{\gtwentysixfourteen}
This object is number 123 in the catalogue of \citet{KobChiSch16}, with the star 2MASS\;J18390969-0603279 of unknown spectral type at the focal point of the IR arc.
It is in projection located in the highly reddened star cluster Stephenson 2 \citep{Ste90, OrtBicBar02}.
It is also close to the centre of the MW bubble MWP1G026137-000346 \citep{SimPovKen12}, which is probably associated with the bow shock and not a separate structure.
The bow shock is clear in our observations (Figure~\ref{fig:detections2}, first row, left panel), with a width of $\sim\ang{;;113}$.
To the south of the bow shock a point source is detected (not the driving star), which has been detected by \citet{BruSchTay21} in their VLA follow-up of unidentified \textit{Fermi-LAT} sources (source VLASS J183908.71-060359.6).

Interestingly Figure~\ref{fig:detections2} shows that we predominantly detect one side of the bow shock, with the eastern side undetected although it is just as bright in IR as the northern side.
This impression is artificially enhanced by the relatively strong gradient in background emission across the image from bottom-left to top-right, which we were not able to satisfactorily remove (possibly because of bright point sources to the south).
So, while this apparent one-sided detection is intriguing, it should be confirmed either through a re-analysis of the dataset or from further observations.

An apparently significant peculiar proper motion is measured opposite to the direction of the bow shock, but the parallax error of the \textit{Gaia} source is large and so our method is likely unreliable.
For this reason we do not plot the proper-motion vector.
Unfortunately little is known about the driving star of the bow shock or its stellar wind.
Our detection in radio implies that this is a true bow shock and not a bow wave because there is enhanced gas emission as well as dust emission.

The spectral index map and its uncertainty, for the bow shock and the point source are shown in the second and third panels of the first row of Figure~\ref{fig:detections2}, respectively.
The point source has a negative spectral index, typical of active galactic nuclei, while the bow shock shows nominally a negative spectral index but with very large uncertainty.
We cannot draw any conclusions about the nature of the emission (thermal or non-thermal) because the uncertainties are too large.

The spatial resolution of the Effelsberg data  is comparable to the long axis of the bow-shock emission seen with the VLA and is therefore sensitive to larger-scale diffuse emission (fourth panel in first row of Figure~\ref{fig:detections2}).
Any emission from the bow shock is diluted by diffuse emission within the large beam.
There are bright point sources to the south that are also contributing significantly to the emission at the position of the bow shock in the Effelsberg map (see Figure~\ref{fig:eff-bow} for a larger-scale view).
The spectral index in the Effelsberg map at the position of the bow shock is consistent with thermal emission.

\subsubsection{\gthirtynine}
This candidate bow shock, entry 214 in the catalogue of \citet{KobChiSch16}, is driven by the star 2MASS\,J19031363+0535230, of unknown spectral type.
There is no \textit{Gaia} source associated with it, and no proper motion information is available.
The region is catalogued as an ultra-compact H~\textsc{ii} region by \citet{BroNymMay96}, id [BNM96] 039.251-0.072.
In \textit{WISE} 22\,$\mu$m images it appears as a small arc around the star, surrounded by a horseshoe-shaped nebula at 12\,$\mu$m, presumably the compact H~\textsc{ii} region with diameter $\approx45^{\prime\prime}$.
The morphology that we detect in the VLA image shown in Figure~\ref{fig:detections2} (fifth row, first panel) matches closely the larger horseshoe-shaped nebula (with a width of $\sim\ang{;;38}$) and we can conclude that we detect the H~\textsc{ii} region.
There is a hint of a double-shelled structure in the VLA image, but the spatial resolution is insufficient to clearly distinguish any putative bow shock from the H~\textsc{ii} region.
This could therefore be a candidate for either a bow shock or for an IR arc around a star moving subsonically within its H~\textsc{ii} region in a dense ISM \citep{MacHawGva16}.

A spectral-index map and its error are plotted in the second and third panels of Figure~\ref{fig:detections2} (fifth row), respectively.
the spectrum is almost flat ($-0.25\lesssim\alpha\lesssim0$) where the emission is bright, indicating that the VLA emission is dominated by thermal emission, while any potential non-thermal emission is too weak to detect.
The nebula is a point source for Effelsberg (fifth row, fourth panel, Figure~\ref{fig:detections2}).

\subsubsection{\gfortynine}
Entry 264 in the catalogue of \citet{KobChiSch16} is a semi-circular arc of mid-IR emission around the star 2MASS\,J19232226+1448087, of unknown spectral type.
In \textit{Gaia} DR3 this star has unknown distance (negative parallax) and so its significant proper motion cannot be converted into a peculiar velocity with respect to its surroundings.
Like \gthirtynine, the putative bow shock is surrounded by a larger ring of IR emission indicating an H~\textsc{ii} region with a somewhat cometary shape.
The H~\textsc{ii} region is catalogued as [WAM82] 049.704-0.172 in \citet{WinAltMez82} (also possibly the radio source WSRTGP\;1921+1442 mentioned in the Westerbork Synthesis Radio Telescope \SI{327}{\MHz} survey; \citealt{TayGosCol96}), and in IR it is somewhat reminiscent of RCW\,120 \citep{ZavPomDeh07}.

The radio map is shown in the sixth row of Figure~\ref{fig:detections2}, where it appears that we detect both the inner arc and an outer incomplete ring of emission in the VLA intensity map (first panel).
The outer edge of the radio emission lines up well with the bright outer arc of emission from \textit{WISE} 12\,$\mu$m, suggesting that we are detecting emission from the photoionised gas and photodissociation region within the H~\textsc{ii} region, as well as the bow shock.

The second and third panels of Figure~\ref{fig:detections2} (sixth row) show the spectral index map and its uncertainty, respectively.
Nominally most of the region has a strongly negative spectral index, although the $1\sigma$  uncertainties are relatively large over most of the emitting region and do not allow us to draw a decisive conclusion.
The uncertainties are smallest at the inner-arc, and the non-thermal spectral index derived from this region is intriguing and should be followed up with further observations at different frequencies.
Figure~\ref{fig:detections2} (sixth row, fourth panel) shows that we detect the region clearly with Effelsberg, but the nebula is too small to detect any substructure and the emission is most likely dominated by the H~\textsc{ii} region.
The spectral index derived from Effelsberg data is flat, consistent with thermal emission.

\subsubsection{\bdforty and its bow shock \bdfortybs}
This source has been detected at a number of radio frequencies \citep{BenRomMar10, BenDelHal21} and was reported on extensively in our previous work \citepalias{MouMacCar22}, where we also detected radio emission from the bow shock.
The negative spectral index of our measurements support the previous detection of non-thermal emission from this bow shock.
It is a largely extended bow shock around a massive O supergiant that has a large peculiar proper motion to the north.
For completeness, the VLA intensity map, spectral index map, spectral-index uncertainty map and Effelsberg intensity map are shown in the second row of Figure~\ref{fig:detections2}, and we refer the reader to \citetalias{MouMacCar22} for further discussion.
The peculiar proper motion re-calculated with \textit{Gaia} DR3 data is plotted on the VLA intensity image, and is consistent with our previous calculation within the uncertainties.

\subsubsection{\bdsixty and its bow shock \bdsixtybs}
The VLA detection of the Bubble Nebula, NGC\,7635, was reported on extensively in our previous work \citepalias{MouMacCar22}.
It is a large bow shock around a massive O giant/supergiant that has a peculiar proper motion to the north.
We detected both thermal and non-thermal emission from the nebula.
The radio detection is reproduced here for completeness in the fourth row of Figure~\ref{fig:detections2},
and we refer the reader to \citetalias{MouMacCar22} for further discussion.

\subsection{Marginal cases}\label{sec:marginal}

\begin{figure*}
    \centering
    \includegraphics[height=0.25\textwidth]{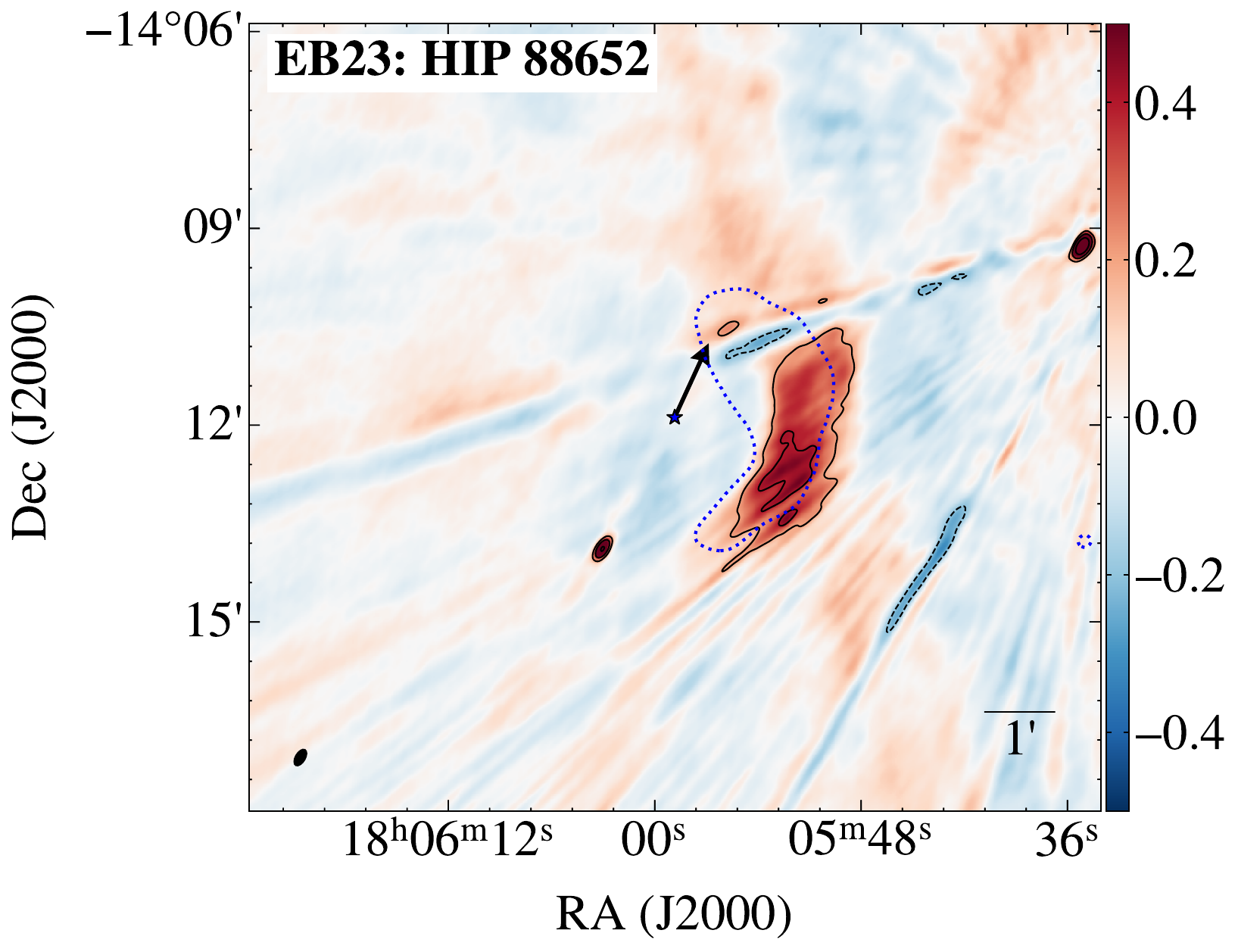}
    \includegraphics[height=0.25\textwidth]{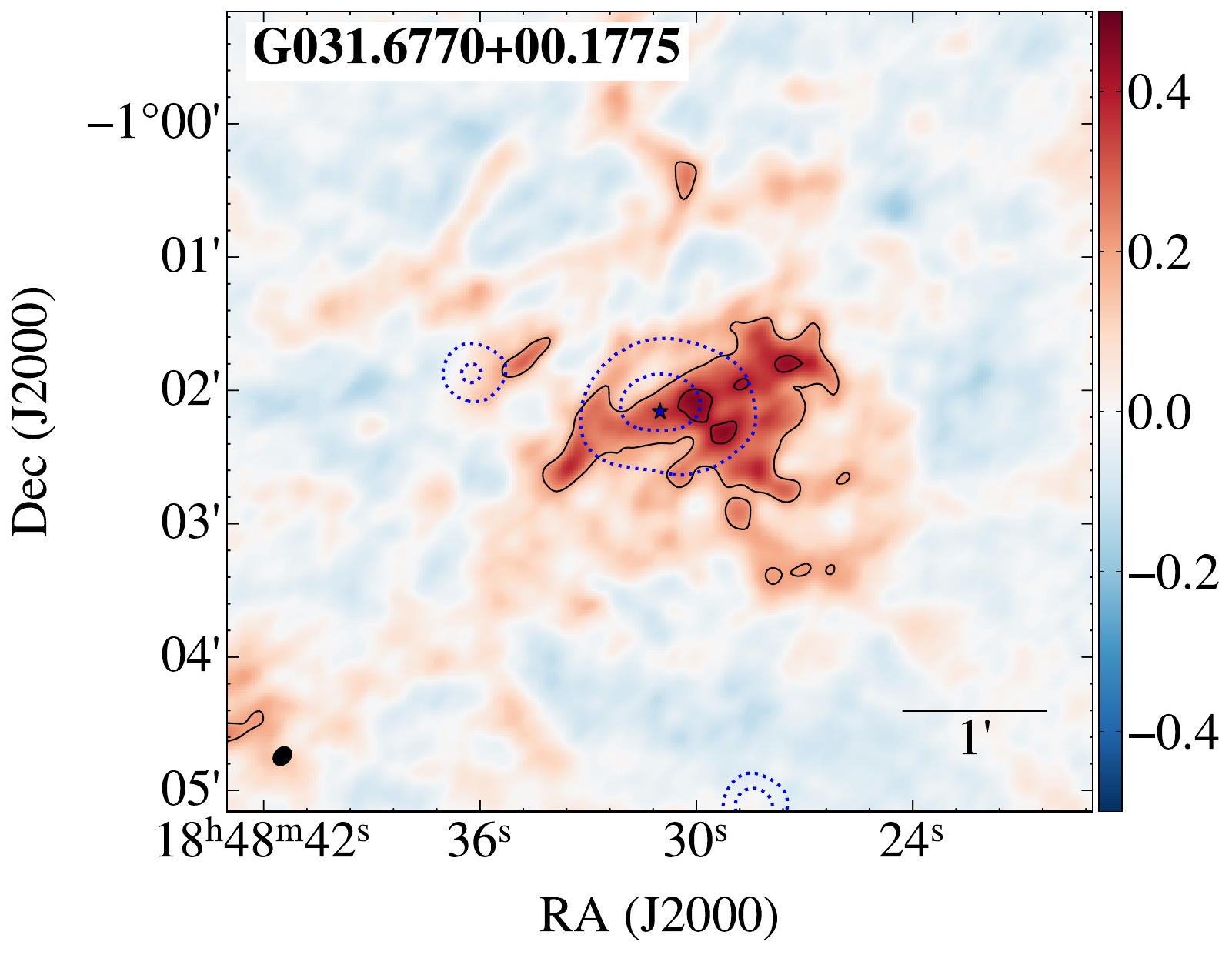}
    \includegraphics[height=0.25\textwidth]{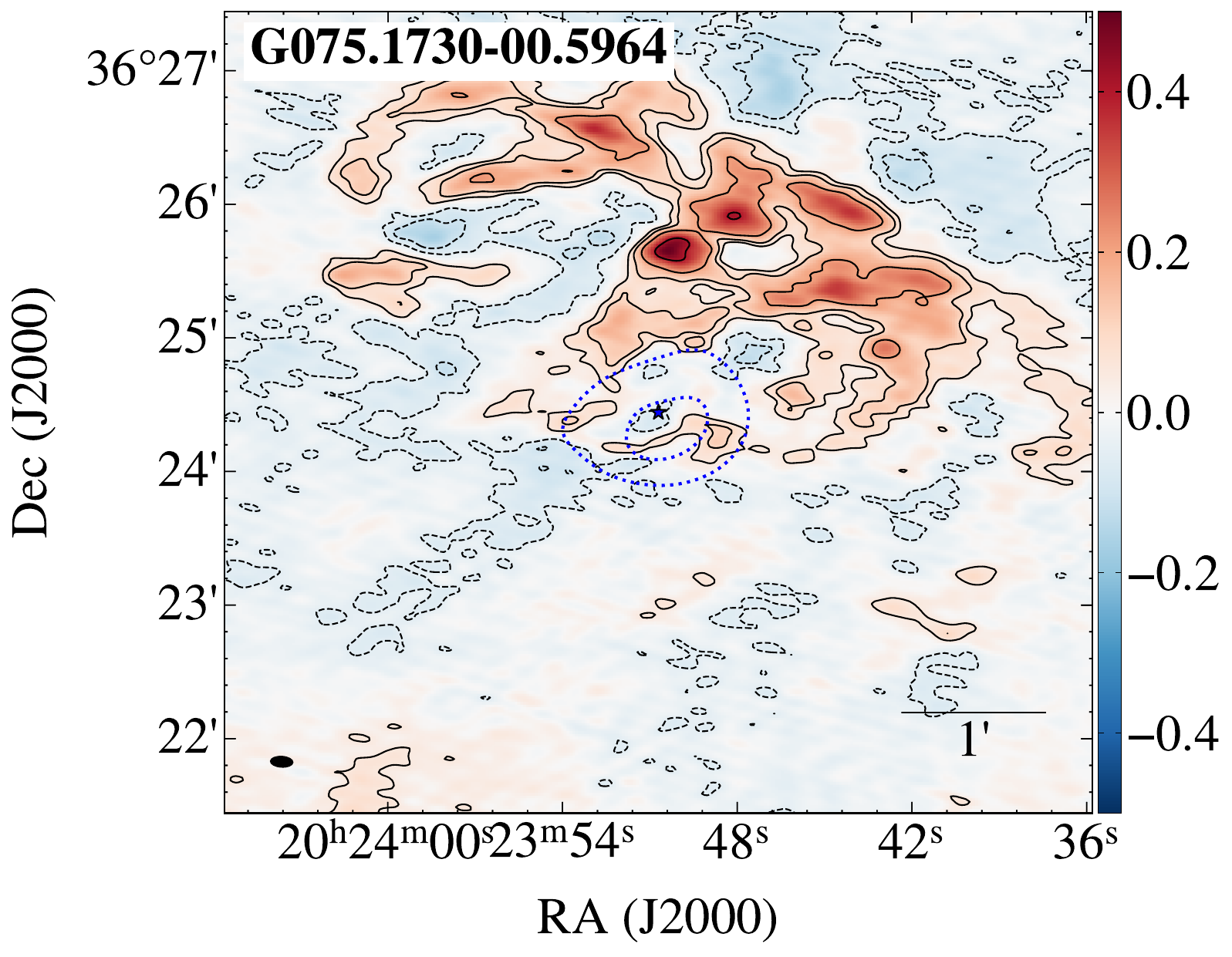}
    \includegraphics[height=0.25\textwidth]{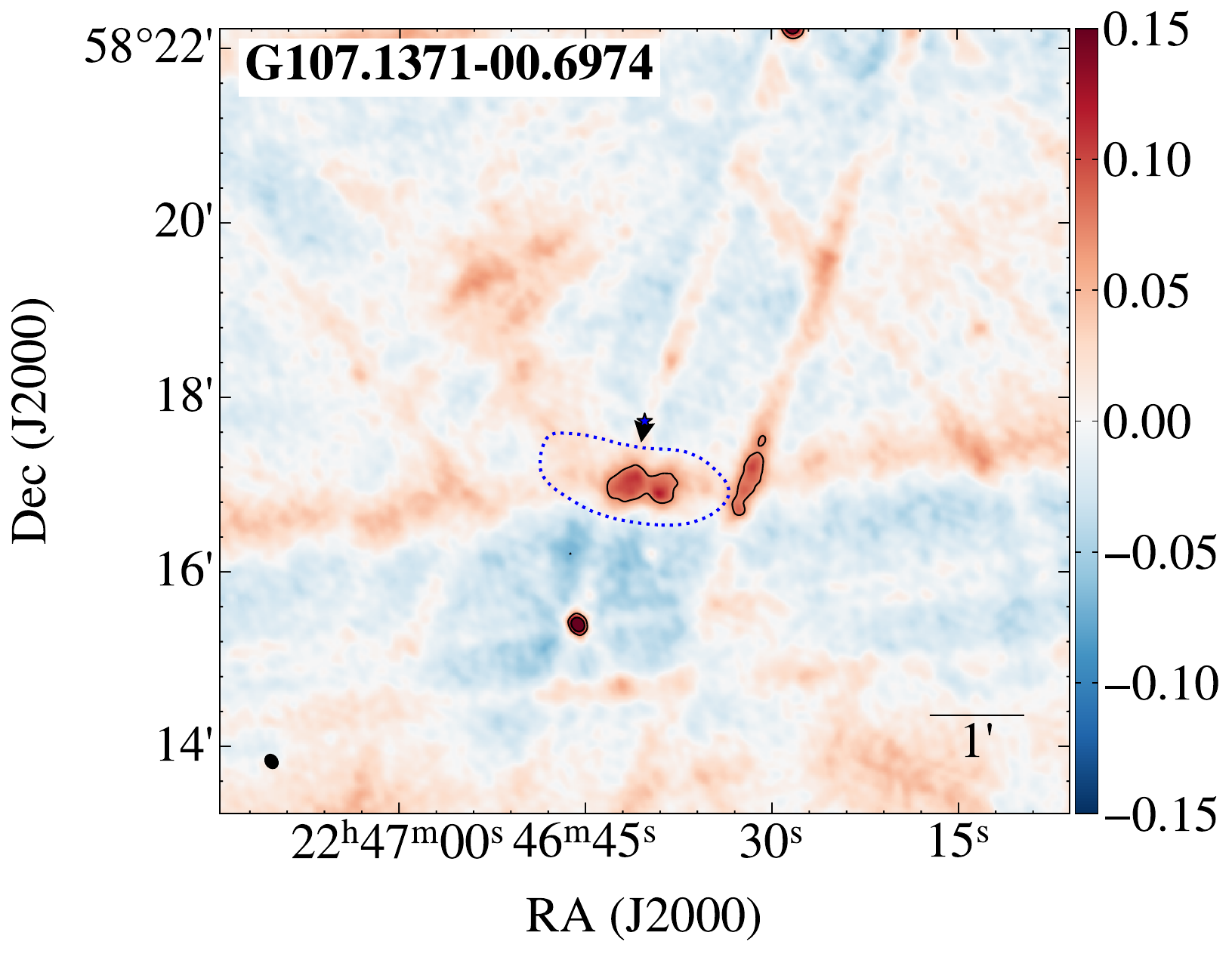}
    \includegraphics[height=0.25\textwidth]{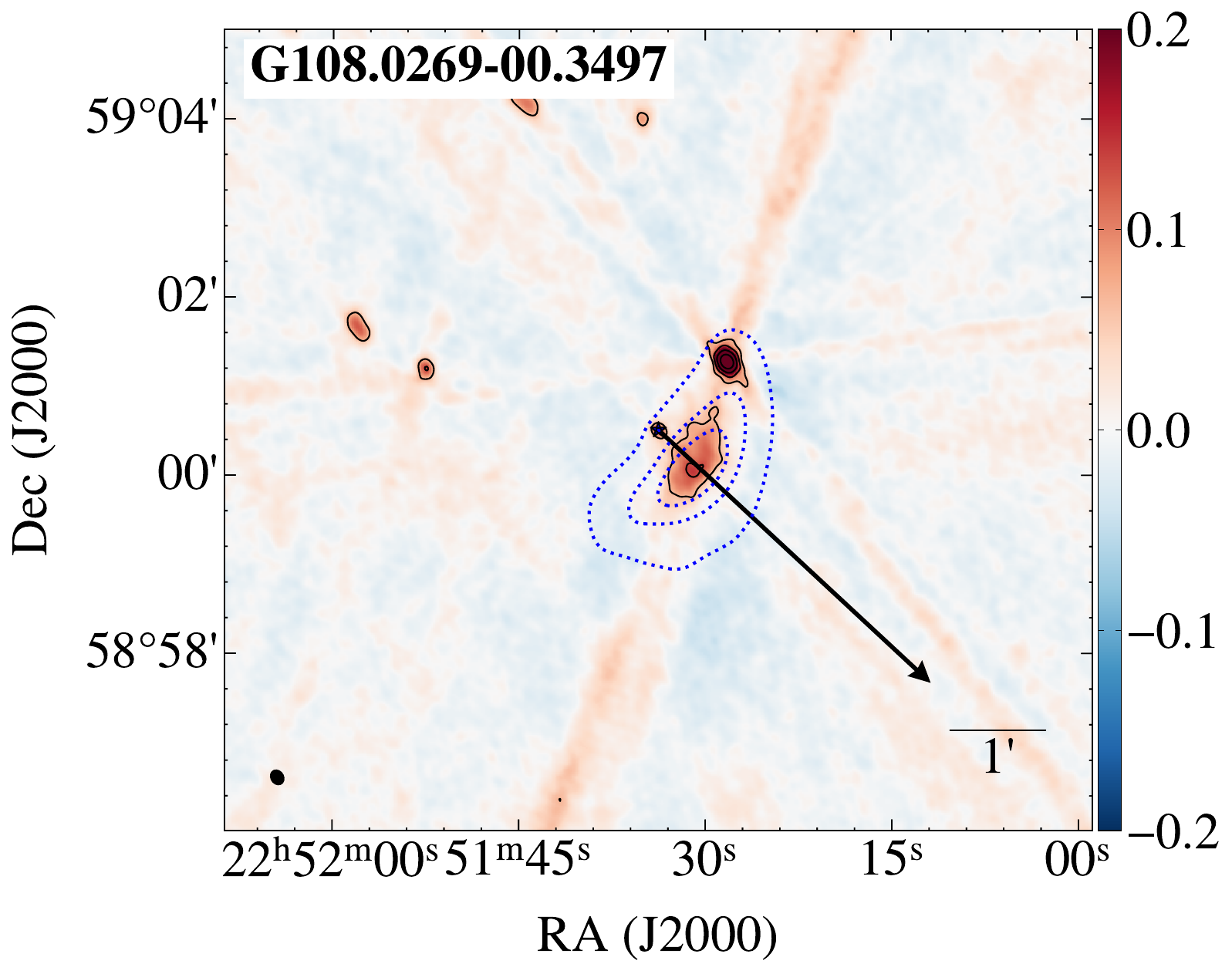}
    \caption{Intensity plots of the targets listed in Table~\ref{tab:targets} for which there is a hint of possible emission but not enough to claim a detection.
    The colour bar values are in mJy\,beam$^{-1}$, with negative values shown in blue.
    In the bottom right a line that corresponds to \ang{;1;} is shown, while in the bottom left the cross-section of the synthetic beam's FWHM can be seen.
    Contours of the most interesting features from the WISE \SI{22}{\um} counterpart are overlaid with blue lines.
    Black contours are radio emission at levels $[-10,-5,5,10,20,40,80,160]\,\sigma$, with negative contours using a dashed line.
    }
    \label{fig:quasi-detections2}
\end{figure*}

For five of our sources there is some indication of radio emission, but not sufficient that we would claim a detection of a bow shock.
The VLA intensity map for each of them is shown in Figure~\ref{fig:quasi-detections2} and each source is discussed in turn below.
As is evident from Figure~\ref{fig:quasi-detections2}, in three of these cases (\hdsixteen, \gonezeroseven and \gonezeroeight) we measure faint emission that is not sufficiently above the noise level of the map to claim a detection, and further investigations with deeper surveys are strongly recommended.
The other two cases show weak evidence for emission with position and morphology consistent with the IR arc, but this is overlaid on brighter diffuse emission from a surrounding H~\textsc{ii} region and so cannot be confirmed.
Further higher-sensitivity observations with higher spatial resolution may be able to disentangle the two components and confirm a detection.

\subsubsection{\hdsixteen}
  This star (aka BD-14$^\circ$\,4880 and HD\;165319) was identified as driving a bow shock in the E-BOSS catalogue \citep{PerBenBro12}, and consists of a 5$^\prime$-long arc of mid-IR emission to the W of the star.
  It is surrounded by a H~\textsc{ii} region with various names including [LPH96] 015.128+3.310 \citep{LocPisHow96}.
  The bow shock and H~\textsc{ii} region are also detected by MSX at 8.3\,$\mu$m and in H$\alpha$ nebular emission \citep{GvaBom08}.
  \citet{PerBenIse15} detected an excess of radio emission at the position of the bow shock in NVSS 1.4\,GHz emission, and \citet{VanSaiMoh22} reported a tentative detection in RACS data at 887.5\,MHz.
  Considering their reported diffuse emission at a level of $\approx1$\,mJy\,beam$^{-1}$ and the quoted beamsize, the surface brightness of their tentative detection is very similar to ours.
  We detect emission from the region of the bow shock, nominally above $10\sigma$ at the apex of the bow shock, but it lies at the point of constructive interference of two stripes of noise that make a faint X-shape on the image and so we cannot claim a secure detection.
  The image is heavily affected by imaging artefacts from the bright point source to the WNW of the star, as also reported by \citet{VanSaiMoh22}.
  We measure a small peculiar proper motion of 19\,km\,s$^{-1}$ in a direction consistent with the orientation of the bow shock (Table~\ref{tab:targets}).
  This is clearly a promising target for follow-up observations, given the consistent results at three different observing frequencies with two different telescopes.

\subsubsection{\gthirtyone}
  This bow-shock candidate, entry 159 in the catalogue of \citet{KobChiSch16}, is actually a region of diffuse mid-IR emission, not clearly arc-shaped in WISE emission, and surrounds the star 2MASS\,J18483100-0102094 with unknown spectral type.
  In \textit{Spitzer} 24\,$\mu$m the higher spatial resolution reveals a faint arc of emission at a standoff distance of about $10\arcsec$ to the N of the star.
  The star and bow-shock candidate are at the centre of  the H~\textsc{ii} region  HRDS G031.677+00.179 \citep{AndBanBal11}.
  We clearly detect diffuse emission around the star in Figure~\ref{fig:quasi-detections2}, presumably free-free emission from the H~\textsc{ii} region.
  There is no clear arc of emission to the NNE that correlates with the IR arc, but the VLA intensity map is dominated by the larger-scale diffuse emission and we cannot claim to detect the bow shock.
  The peculiar proper motion we obtain is unreliable because of the large uncertainties in the \textit{Gaia} parallax  (Table~\ref{tab:targets}).

\subsubsection{\gseventyfive}
  This bow-shock candidate, entry 323 in \citet{KobChiSch16}, is surrounded by the H~\textsc{ii} region [LPH96] 075.167-0.575 \citep{LocPisHow96} and driven by the central star 2MASS\,J20235070+3624265, of undetermined spectral type.
  Table~\ref{tab:targets} shows that we could not measure a reliable peculiar proper motion for the central star because the \textit{Gaia} parallax has very large uncertainties, and so we do not plot the proper motion in Figure~\ref{fig:quasi-detections2}.
  There is a $5\sigma$ contour of weak emission coinciding with the IR arc but, because the emission is largely dominated by the H~\textsc{ii} region, we cannot claim this as a detection.
  Indeed the background emission has many positive and negative $5\sigma$ fluctuations, implying that the region we used to extract the RMS noise level may be underestimating the true noise level of the map.
  We also detect significant diffuse emission that corresponds to the H~\textsc{ii} region, brighter to the N than to the S in IR images, and spatially correlated with the \textit{WISE} 12\,$\mu$m emission.
  The presented dataset does not allow us to conclude whether the star is a runaway nor if there is gaseous emission from the bow shock.

\subsubsection{\gonezeroseven}
  This bow-shock candidate, entry 359 in the catalogue of \citet{KobChiSch16}, is located to the S of the driving star (aka BD+57$^\circ$\,2606 and HD\,215806), reported as an Oe star by \citet{ChiKobSch20}.
  It is located on the periphery of the blister H~\textsc{ii} region [ABB2014] WISE G107.156-00.988 \citep{AndBanBal14} driven by the open cluster NGC\,7380, with gas expansion in the direction opposite to the orientation of the bow shock.
  As such it could be an example of the `interstellar weather vanes' found by \citet{PovBenWhi08} in M17 and RCW\,49, where the orientation of the bow shock traces not the motion of the star, but a bulk flow of gas escaping from a high-pressure region.
  Supporting this possibility, we detect no significant peculiar proper motion ($\approx5$\,km\,s$^{-1}$) for the driving star (Table~\ref{tab:targets}).
  We detect weak emission in the region of the IR arc, shown in Figure~\ref{fig:quasi-detections2}, but again the emission is located at the superposition of noise stripes across the image and so we cannot claim a detection.

\subsubsection{\gonezeroeight}
  Entry 360 in the catalogue of \citet{KobChiSch16}, this bow shock candidate is located to the SW of the driving star BD+58$^\circ$\,2492 (aka HD\,216411), a B0.7Ia supergiant \citep{NegSimdeB24}.
  Figure~\ref{fig:quasi-detections2} shows that we detect faint emission spatially coincident with the bow shock, but it is superimposed on a stripe of noise from a nearby bright point source that enhances the signal artificially above the $5\sigma$ threshold.
  We also measure a large and well-constrained peculiar proper motion of the star of $\approx85$\,km\,s$^{-1}$ in a direction appropriate for the orientation of the bow shock.

\subsection{Non-detections}

Of the 24 targeted sources, 13 were not detected by the VLA and these fields are shown in the Appendix in  Figure~\ref{fig:non-detection-images2}.
Some are clearly affected by bright sources outside the field of view, whereas others simply have no detection down to the RMS noise level of the map, either because we are not sensitive to the spatial scales involved (which should not occur because of our selection criteria) or because the gaseous emission is much weaker than expected based on the bright IR emission from dust.
Discussion of each individual target is in the Appendix.

\begin{table*}[]
    \centering
    \caption{Estimation of $n_e$ based on size and brightness of bow shock (or RMS noise level of the map for non-detections) using Eq.~\ref{eq:ne}.}
    \label{tab:ne}
    \begin{tabular}{l|c c c c c c c}
    Source & $d$ & method & $R_0$  & RMS  & $I_\nu$  & UL? & $n_e$ \\
           & (kpc)  &     & (pc)   & (mJy\,beam$^{-1}$) & (mJy\,beam$^{-1}$) &  &  (cm$^{-3}$) \\
    \hline
    \goneonenine        & 2.27 & GSP-Phot & 0.19 & \num{0.02} & $<0.06$ & yes & $<51.9$ \\
    \hiptwenty          & 1.18 & GSP-Phot & 0.29 & \num{0.01} & $<0.03$ & yes & $<28.9$ \\
    \sersix             & 2.35 & para     & 0.34 & \num{0.2} & $<0.60$ & yes & $<124$ \\
    \gonethreethree     & 2.10 & para     & 0.37 & \num{0.009} & $<0.027$ & yes & $<19.5$ \\
    \gonethreefour      & 1.98 & para     & 0.13 & \num{0.05} & $<0.15$ & yes & $<75.6$ \\
    \gonethreeseven     & 1.09 & GSP-Phot & 0.39 & \num{0.01} & $<0.03$ & yes & $<28.4$ \\
    \gonefiveone        & 1.77 & GSP-Phot & 0.44 & \num{0.02} & $<0.06$ & yes & $<38.7$ \\
    \hiptwentysix       & 1.19 & GSP-Phot & 0.35 & \num{0.004} & $<0.012$ & yes & $<12.9$ \\
    \hipthirtyone       & 1.08 & GSP-Phot & 0.63 & \num{0.015} & $<0.045$ & yes & $<15.4$ \\
    \hdfiftyseven       & 0.71 & GSP-Phot & 0.06 & \num{0.01} & $<0.030$ & yes & $<41.3$ \\
    \zoph               & 0.135 & (2)     & 0.13 & \num{0.018} & $<0.054$ & yes & $<39.7$ \\
    \hdsixteen          & 1.46 & para & 0.67 & \num{0.04} & 0.4 & no & $<43.5$ \\
    \geighteen          & 4.14 & para & 0.30  & 0.5 & 15 & no & $<411$ \\
    \gtwentysixfifytwo  &  n/a & n/a      & n/a & \num{0.5}  & $<1.5$   & yes & n/a  \\
    \gtwentysixfourteen & n/a & n/a & n/a & 0.12 & 1.5 & no & n/a \\
    \gthirtyone         &  n/a & n/a & n/a & \num{0.04} & 0.2 & no & n/a  \\
    \gthirtynine        &  n/a & n/a & n/a & \num{1.5} & 40 & no & n/a  \\
    \gfortynine         &  n/a & n/a & n/a & \num{0.19} & 2.0 & no & n/a  \\
    \gseventyfive       & 2.85 & GSP-Phot & 0.13 & \num{0.01} & 0.05 & no & $<62.3$ \\
    \gseventyeight      & 1.79 & GSP-Phot & 0.22 & \num{0.08} & $<0.24$  & yes & $<60.9$ \\
    \bdforty            & 1.72 & (1) & 1.61  & 0.07 & 1.4 & no  & $<69.5$ \\
    \gonezeroseven      & 2.64 & para & 0.54 & \num{0.014} & 0.07 & no & $<26.7$ \\
    \gonezeroeight      & 3.03 & para & 0.52 & \num{0.013} & 0.13 & no & $<37.0$ \\
    \bdsixty            & 3.0 & (1) & 0.60  & 0.2 & 5.0 & no & $<252$ \\
    \end{tabular}
    \tablebib{
    (1)~\citetalias{MouMacCar22}; (2)~\citet{GreMacKav22}.
}
\tablefoot{
The column `method' refers to how the distance was estimated:
numbers refer to cited references,
`GSP-Phot' refers to \textit{Gaia} DR3 GSP-Phot distance estimate where available,
`para' refers to inverted parallax when the `GSP-Phot' distance is not available and the relative uncertainty on the parallax is $<1/8$,
and `n/a' is quoted where the distance is unknown and no estimate for $R_0$ or $n_e$ was attempted.
The angular estimate for $R_0$ is taken from \citet{KobChiSch16} and \citet{PerBenIse15}, depending on which catalogue the bow-shock candidate is located in, and converted to a linear size using the distance estimate.
The intensity around the apex of the bow shock, $I_\nu$, is measured from the observations, or a $3\sigma$ upper limit is taken for non-detections using the RMS level of the map (column 5).
The estimated electron density, $n_e$, is calculated from Eq.~\ref{eq:ne} and is always an upper limit because part of the emission could be synchrotron and not bremsstrahlung.
}
\end{table*}

\subsection{Limits on electron density from radio emission and upper limits} \label{sec:electrondensity-results}

Following the method outlined in Section~\ref{sec:electrondensity}, we can obtain upper limits on the electron density, $n_e$, for the detections and upper limits.
The results of this estimate for our 24 sources are shown in Table~\ref{tab:ne}, assuming an observing frequency $\nu=6$\,GHz.
For the three well-studied bow shocks in our sample, around the stars \zoph, \bdforty{} and \bdsixty, we can compare our estimated $n_e$ to literature measurements.

\begin{figure*}
    \centering
    \includegraphics[width=\textwidth]{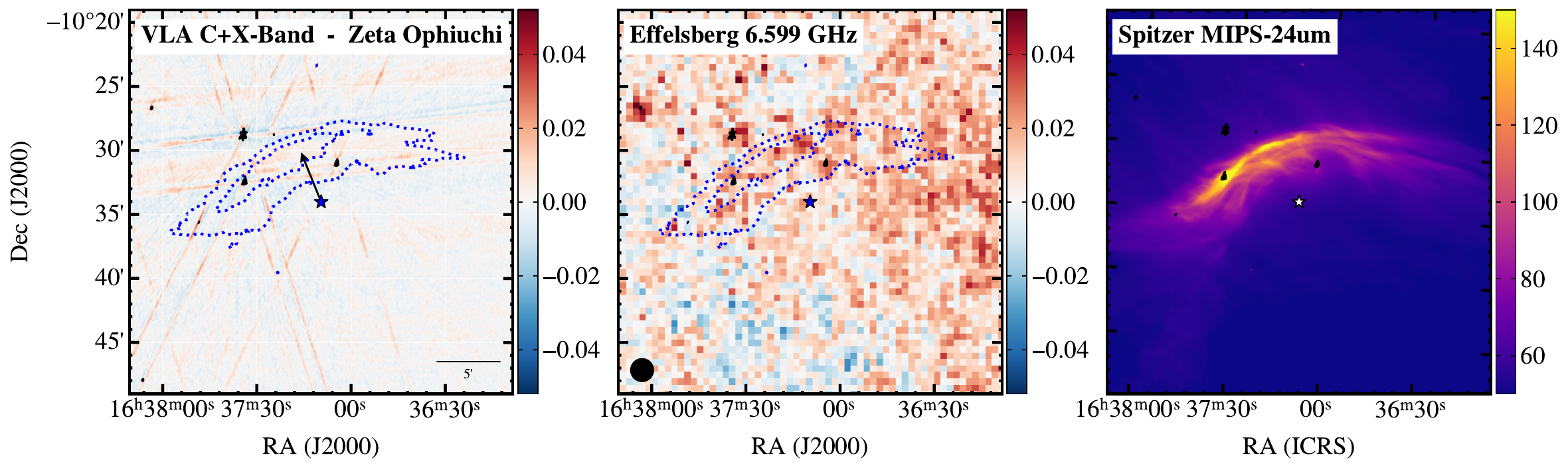}
    \caption{
    Comparison of observations of the bow shock of \zoph{} with the VLA (left), Effelsberg (centre), and the mid-IR image from \textit{Spitzer Space Telescope} MIPS 24\,$\mu$m.
    All three images are plotted in units of megaJansky per steradian.
    The VLA and Effelsberg beams are plotted at the lower-left of the respective panels.
    For the VLA and Effelsberg plots, IR emission contours at levels 70 and 100\,MJy\,sr$^{-1}$ are overplotted in dotted blue lines, and for Effelsberg and \textit{Spitzer} plots the VLA intensity contours are overlaid in black, showing only the detected point sources.
    }
    \label{fig:zeta-oph}  
\end{figure*}

\paragraph{\zoph:} \citet{GvaLanMac12} estimated $n_e\approx3$\,cm$^{-3}$ for the ISM upstream of the bow shock, and our upper limit for the shocked layer ($n_e\lesssim39.7$\,cm$^{-3}$) corresponds to a compression factor $\leq13$, although it should be noted that for this very extended bow shock the VLA may be missing flux from large scales, weakening this constraint.
\citet{GreMacKav22} simulated the bow shock of \zoph and made synthetic 6\,GHz bremsstrahlung emission maps, predicting a peak brightness of $I_\nu\approx0.035$\,MJy\,sr$^{-1}$, corresponding to $0.10$\,mJy\,beam$^{-1}$ for the VLA beam.
This is twice the $3\sigma$ upper limit from Table~\ref{tab:ne}, so either the ISM density considered in \citet{GreMacKav22} was significantly too large, or we are losing a lot of flux in the VLA image.
To test this, \zoph is the only bow shock in our sample that should be spatially resolved with Effelsberg, and so we can perform the same analysis with Effelsberg data.
Using maps at 4.951\,GHz and 6.599\,GHz, with RMS noise levels of 3.8\,mJy\,beam$^{-1}$ and 3.5\,mJy\,beam$^{-1}$, beamsizes of $141\arcsec$ and $110\arcsec$, respectively, we obtain $3\sigma$ upper limits of $n_e(\nu=4.951\,\mathrm{GHz})\lesssim44.7$\,cm$^{-3}$ and $n_e(\nu=6.599\,\mathrm{GHz})\lesssim54.6$\,cm$^{-3}$.
The constraints are similar to (but higher than) the VLA limit, and are significantly more robust because there is no loss of large-scale flux.
A comparison of the VLA and Effelsberg data is shown in Figure~\ref{fig:zeta-oph}, where we plot VLA, Effelsberg, and \textit{Spitzer MIPS} 24\,$\mu$m data all in units of MJy\,sr$^{-1}$.
It is notable that the mid-IR emission is at least $10^4\times$ more luminous than the radio emission.

\paragraph{\bdsixty:} Our estimate of $n_e\lesssim250$\,cm$^{-3}$ is larger than estimates of $n_e\approx100-200$\,cm$^{-3}$ from optical nebular spectroscopy \citep{EstBesMor16}.
The similarity of these numbers nevertheless implies that thermal emission should be contributing measurably to the total emission at 6\,GHz and should start to dominate at higher frequency.
It should be noted that the crude assumptions of constant $n_e$ and the simple geometry likely introduce uncertainties of at least a factor of roughly two in the estimates of $n_e$, so we cannot draw any clear conclusions about the relative contributions of thermal and non-thermal emission to the flux.

\paragraph{\bdforty:} The gas density in the pre-shock ISM was estimated by \citet{ComPas07} to be $n_e\approx6$\,cm$^{-3}$, and so our upper limit of $n_e\lesssim69.5$\,cm$^{-3}$ implies a compression factor of $\lesssim12$.
We estimate the space velocity of the star to be $v_\star\approx45$\,km\,s$^{-1}$ (Table~\ref{tab:targets}), which would give a Mach number of $\approx4$ and compression factor of $\approx16$ for an isothermal shock in the photoionised ISM.
As noted for \bdsixty, these crude estimates have rather large and poorly quantified uncertainties, but imply that thermal emission should be contributing much of the emission at and above 6\,GHz.

In both cases of \bdforty{} and \bdsixty{}, further observations at lower and higher frequencies should be performed to measure the transition from a spectrum dominated by synchrotron emission (at low $\nu$) to one dominated by thermal emission (at high $\nu$).
The lower frequencies have been measured by \citet{BenDelHal21} for \bdforty, but higher frequency measurements would be valuable to measure the thermal contribution.
A thermal contribution to the flux from the bow shock would reduce the large energy-requirements for high-energy particles noted in \citetalias{MouMacCar22}.
They assumed that all of the emission was synchrotron and found that it was difficult to inject enough energy into the region to explain the emission, particularly for \bdsixty{} where both newly accelerated particles and compressed interstellar cosmic-ray electrons struggled to explain the level of emission.

It should be noted that the H~\textsc{ii} regions around \bdsixty{} and \bdforty{} should contribute to the emission measure in addition to thermal and non-thermal emission from the bow shock, although the VLA observations may not be sensitive to any large-scale uniform emission.
Indeed Fig.~\ref{fig:detections2} shows no large-scale diffuse emission for either of these sources, indicating that The VLA has resolved out the H~\textsc{ii} region and it does not contribute to the estimated emission measure.

For the other sources we obtain the upper limits $n_e\in[12-125]$\,cm$^{-3}$, which is appropriate for shock-compressed gas in the diffuse ISM, considering compression factors $\approx10$ and bearing in mind that bow shocks are more likely to be detected for stars moving in higher density environments.
The exception is \geighteen, which is located at the edge of a young star cluster in a massive star-forming region, and so the measured $n_e\lesssim411$\,cm$^{-3}$ is much larger than for the stars in the diffuse ISM.
Alternatively it could be an indication of strong non-thermal emission from the bow shock.
Considering also the size of the bow shock, $R_0$, we estimate upper limits on the emission measure, $EM=\int n_e^2 d\ell$, in the range 57 to 5200\,cm$^{-6}$\,pc, with typical values in the range 100 to 500\,cm$^{-6}$\,pc.

We can compare our upper limits with those derived from the measured $R_0$ and $v_\star$, together with estimated mass-loss rates and wind velocities in Table~\ref{tab:app:mdot}.
Re-arranging Equation~\ref{eqn:standoff} gives the unshocked ISM density, and the associated electron number densities, $n_{e,0}$, are quoted in Table~\ref{tab:app:mdot}.
In most cases $n_{e,0}< n_e$ with the notable exception of \hdsixteen, for which we obtain a large $n_{e,0}=120.5$\,cm$^{-3}$ on account of the large mass-loss rate ($\dot{M}=5.3\times10^{-6}\,\mathrm{M}_\odot\,\mathrm{yr}^{-1}$) suggested by \citet{GvaBom08}.
O9.5I supergiants have a large range of measured $\dot{M}$, however, and if we use the lowest value from \citet{MokDeKVin07} of $\dot{M}=7\times10^{-7}\,\mathrm{M}_\odot\,\mathrm{yr}^{-1}$ then the ISM density is reduced to $n_{e,0}=15.9$\,cm$^{-3}$.
This is still quite large, but no longer in conflict with the radio-derived upper limit on $n_e$, and consistent with the density estimate of the surrounding H~\textsc{ii} region by \citet{GvaBom08}.

\section{Discussion} \label{sec:discussion}
This work constitutes the largest targeted radio survey of bow shocks to date, and the detection rate of 25\% for clear detections, and 46\% when also including possible detections, demonstrates the success of our targeted approach.
The only research on a similar scale was by \citet{VanSaiMoh22}, who used the publicly released Rapid ASKAP Continuum Survey \citep{McCHalLen20} data to search for radio emission from 50 IR-selected bow shocks, finding 3 clear detections and 3 tentative detections, a success rate of 6\% (12\%) for clear (tentative) detections.
The lower success rate of \citet{VanSaiMoh22} may be attributed to the fact that their dataset was a very wide-field survey rather than targeted deep radio observations.
Using a less sensitive and lower resolution dataset (the NVSS), \citet{PerBenIse15} could find associations of radio emission with a number of bow shocks, some of which were subsequently confirmed by \citet{VanSaiMoh22}.
This highlights the necessity of deep radio observations to detect emission from bow shocks.

Despite this success, of our six detections only 3 are clearly bow shocks whereas the other three appear to have strong contaminating emission from the surrounding H~\textsc{ii} region, which limits the degree to which we can interpret our results.
Similarly for the tentative and possible detections that require follow-up confirmation, two of the five sources have significant contamination from surrounding diffuse emission.
Only approximately half of our firm and possible detections appear as a clean bow-shock structure in radio emission, whereas all are clearly arc-shaped bow shocks in mid-IR imaging.
Similar to the survey of \citet{VanSaiMoh22}, we find tantalising evidence that slightly deeper observations would give clear detections of even more bow shocks.
Our results should be used to plan future observations, including consideration of the locations of bright background sources (particularly for \hdsixteen, \gonezeroseven and \gonezeroeight).
Assuming these are background active galactic nuclei with steep spectra, they could be even more problematic for imaging at lower frequencies where non-thermal emission from the bow-shock candidates would be stronger.

If we do not detect gas emission from a bow-shock candidate, it could indicate that the IR emission comes from pile-up of dust grains that are dynamically decoupled from the ISM by the radiation pressure of the star - a dust wave \citep{HenArt19a} in which the overdensity of dust is not coincident with a gas overdensity.
Alternatively it could merely indicate insufficient sensitivity to the gaseous emission in radio.
For the non-detections, the $3\sigma$ upper limits on the emission can be converted to upper limits on $n_e$ in the shocked layer (with some symmetry assumptions) in the range $\lesssim[12-125]$\,cm$^{-3}$, with a few tens being typical.
For an adiabatic shock in a low-density environment ($n_e\sim[0.1-1]$\,cm$^{-3}$) this is not very constraining and deeper observations are required (or analysis techniques that better remove the striping artefacts from bright sources within and outside the field of view).
For an isothermal shock, typical of denser environments with short cooling timescales, we expect a compression factor of approximately ten for runaway stars with the typical space velocity $v_\star\approx30$\,km\,s$^{-1}$ for a runaway star, meaning that our upper limits constrain the ISM density to $n_e\lesssim[1.2-12]$\,cm$^{-3}$.
The upper limits are therefore not unreasonable for any of the undetected sources and we cannot conclude that gas-dust decoupling is taking place.

For \zoph{} our upper limits on the emission with the VLA and Effelsberg are already close to being in conflict with predictions from magnetohydrodynamic simulations \citep{GreMacKav22}, and so a modest increase in sensitivity of a factor of three should be sufficient to either detect the bow shock or rule out much of the parameter space in wind power and ISM density considered by \citet{GreMacKav22}.
These results underline how much more luminous bow shocks are in mid-IR dust emission than in tracers of gas emission, as noted by predictions from simulations \citep{MeyMacLan14, MacHawGva16, AcrSteHar16}.
Our upper limit for radio emission from \zoph{} shows that the mid-IR emission is at least $10^4\times$ more luminous than at 6\,GHz.

Nearly all of our detections and possible detections show strongly negative spectral index, $\alpha$, in the VLA analysis, but all have flat spectra consistent with optically thin bremsstrahlung at the source position in the Effelsberg data.
Furthermore, the uncertainties on the VLA-derived $\alpha$ are large for most sources, except for \bdforty{}, \bdsixty{} and the very bright sources \geighteen and \gthirtynine, both of which show $\alpha$ consistent (at least marginally) with thermal emission from the surrounding H~\textsc{ii} region.
\gfortynine shows good evidence for non-thermal emission from the bow shock, which should be followed up with further observations, although for this object our lack of knowledge about the driving star limits what can be learned from the nebula.

\citet{Gre22} noted that interferometric observations can give spuriously negative spectral indices for faint and diffuse sources and so, given the relatively large uncertainties in the VLA-derived spectral indices for the fainter sources, we do not consider them reliable enough to make any claims about the emission mechanism.
We attempted combining the single-dish and interferometric datasets using the CASA task `feather' \citep{RauNaiBra19}, as was done in \citetalias{MouMacCar22} for \bdforty{} and \bdsixty{}, but the outcome was less convincing for these other sources because the resulting spectral index obtained was strongly dependent on the choice of weighting.
Attempting to plot the spectrum in the 4-12\,GHz range using the VLA and Effelsberg sub-bands was not attempted for these sources.
With the current dataset we cannot claim to have convincingly discovered any more non-thermal-emitting bow shocks, but nor can we exclude that the bow shocks have non-thermal emission.
Further observations, ideally at similar sensitivity and spatial resolution but at lower (or higher) frequency, are required to determine a robust spectral index for the faint diffuse emission.
It may be possible to detect polarised emission, at least for the bright bow shocks and nebulae in our sample, which would be a clear signature of synchrotron emission, independent of the spectral-index measurement \citep[e.g.][]{BenDelHal21}.

\citet{KobChi22} also measured the 2D space velocity of massive stars with bow shocks.
Of our 24 targets, only \bdforty{} is contained in their table 1, for which our peculiar velocity is almost identical.
Their table 3 contains numbered nebulae but it is not clear how to relate these numbers to our sources and so we could not make a comparison.
Similarly, \citet{CarRibPar23} and \citet{CarBenPar25} studied kinematics of stars driving bow shocks.
There are four common targets with our survey, \hdfiftyseven, \hdsixteen, \bdforty and \bdsixty, and for all of these we find consistent peculiar velocities within $1\sigma$ uncertainties by comparing with their $V_\textsc{pec}^\mathrm{2D}$ values.

Similar to \citet{KobChi22} and \citet{CarBenPar25}, we find the majority of our stars are moving slowly through their environment.
Out of 18 stars with well-determined $v_\star$, we find only 3 with $v_\star \geq 30$\,km\,s$^{-1}$, 6 more with $15 \,\mathrm{km\,s}^{-1} \leq v_\star < 30 \,\mathrm{km\,s}^{-1}$, and 9 with  $v_\star < 15 \,\mathrm{km\,s}^{-1}$.
This means we have only 3 stars satisfying the classical definition of a runaway star ($v_\star \geq 30$\,km\,s$^{-1}$), with 15 being classified as walkaway stars ($v_\star < 30 \,\mathrm{km\,s}^{-1}$), at least when considering only the motion in the plane of the sky.

\section{Conclusions} \label{sec:conclusions}
We have carried out a survey of 24 IR-selected bow shocks in the radio C  and X bands with the VLA, following up on ten of these targets with Effelsberg single-dish C band observations.
We clearly detected arc-shaped emission from six of the 24 targets. In three of these sources, the interpretation is somewhat complicated by the surrounding diffuse radio emission (presumably) from the H~\textsc{ii} region that the bow shocks are embedded in.
For five more sources, we observed evidence of emission that could be a low-significance detection of the bow shock, but noise in the map and imaging artefacts from bright point sources prevents a convincing detection.
In two of these cases there is significant emission from a surrounding H~\textsc{ii} region that further obscures the picture.

Excluding the bright sources \bdforty{} and \bdsixty{} with synchrotron spectra reported on in \citetalias{MouMacCar22}, two of the bright sources (\geighteen and \gthirtynine) have a relatively flat spectral index that may be consistent with thermal emission. \gfortynine may be a non-thermal emitter, and \gtwentysixfourteen has such a large uncertainty on the spectral index map that we cannot draw any conclusions.
Therefore, we do not add to the three confirmed non-thermal-emitting bow shocks that have been reported in the literature, but we identify strong candidates.

Interpreting the entirety of the radio emission as thermal bremsstrahlung emission in the optically thin limit, we estimated upper limits on the gas density within the shocked-ISM layer of the bow shock, finding that all upper limits are consistent with expectations for shock-compressed bow shocks of stars moving through the diffuse ISM in the Galactic Plane.
For the case of \bdsixty, the well-constrained gas density obtained from optical spectral lines is lower than our upper limits, which is consistent with the conclusion of \citetalias{MouMacCar22} that a significant part of the emission is non-thermal.
For this case it appears that both thermal and non-thermal emission contribute at a similar level to the emission at $\approx6$\,GHz, so broader spectral coverage should be able to constrain both components.

Our attempts to detect the bow shock of \zoph{} were unsuccessful both with the VLA and Effelsberg observations.
Comparison of the derived upper limits with simulation results suggests that improved sensitivity by a factor of three with Effelsberg should yield a detection or would very strongly constrain possible models for the bow shock.

Follow-up observations of our confirmed and possible detections at both higher and lower frequencies are strongly encouraged.
These would allow the contributions of thermal and non-thermal emission from the sources to be disentangled and give useful constraints on the gas density and the population of relativistic electrons.

\section*{Data availability}

FITS files used to generate Figures \ref{fig:detections2}, \ref{fig:quasi-detections2} \ref{fig:zeta-oph} \ref{fig:non-detection-images2} and \ref{fig:eff-bow} are available at the CDS via anonymous ftp to cdsarc.u-strasbg.fr (130.79.128.5) or via http://cdsweb.u-strasbg.fr/cgi-bin/qcat?J/A+A/.

\begin{acknowledgements}
JM acknowledges support from a Royal Society-Science Foundation Ireland University Research Fellowship. NC gratefully acknowledges funding from the Deutsche Forschungsgemeinschaft (DFG) – CA 2551/1-2. PJH acknowledges support from Universidad Nacional Aut{\'o}noma de M{\'e}xico Postdoctoral Program (POSDOC). YG was supported by the National Natural Science Foundation of China (Grant No. 12427901), the Strategic Priority Research Program of the Chinese Academy of Sciences (Grant No. XDB0800301), and the Ministry of Science and Technology of China under the National Key R\&D Program (Grant No. 2023YFA1608200). 
This publication results from research conducted with the financial support of Taighde \'Eireann - Research Ireland under Grant numbers 14/RS-URF/3219, 17/RS-EA/3468, 20/RS-URF-R/3712, 22/RS-EA/3810, IRCLA\textbackslash 2017\textbackslash 83.
M.R.R is a Jansky Fellow of the NSF National Radio Astronomy Observatory, USA. 
This work is based on observations with the National Science Foundation’s (NSF) Karl G. Jansky Very Large Array (VLA), operated by the National Radio Astronomy Observatory (Project ID: 19B-105; PI: M.~Moutzouri).
The National Radio Astronomy Observatory is a facility of the National Science Foundation operated under cooperative agreement by Associated Universities, Inc.
This research made use of the Common Astronomy Software Applications \citep[CASA,][]{2007ASPC..376..127M}.
This work is based on observations with the 100 m telescope of the MPIfR (Max-Planck-Institut f{\"u}r Radioastronomie) at Effelsberg (Project ID: 86-20; PI: M.~Moutzouri).
We thank the operations team at the Effelsberg 100 m telescope for their assistance with our observations.
This research has made use of the SIMBAD database, operated at CDS, Strasbourg, France \citep{WenOchEgr00}.
This research has made use of the Astrophysics Data System, funded by NASA under Cooperative Agreement 80NSSC25M7105.
This research made use of the python packages APLpy, \citep{2012ascl.soft08017R}, Astropy \citep{astropy:2018},  Numpy \citep{HarMilVan20}, matplotlib \citep{Hun07}.
\end{acknowledgements}

\bibliographystyle{aa_url}
\bibliography{references}

\begin{appendix}

\section{Targets} \label{sec:app:targets}

Table~\ref{tab:targets} shows a list of the 24 bow-shock candidates targeted with VLA observations, including literature information on the driving star's name/ID, spectral type, parallax and peculiar proper motion, where these could be reliably estimated, as discussed in Section~\ref{sec:methods}.

\begin{sidewaystable*}
\caption{List of VLA targets and their driving star, sorted according to right ascension.} \label{tab:targets}
\centering
    \vspace{-8pt}
\begin{tabular}{l p{2.25cm} rrlc|rr|rr|rl}
\toprule
\midrule
Bow Shock & Driving Star
& R.A.$_\text{J2000}$  
& Dec.$_\text{J2000}$	 
& spec. type 
& Det?
& \multicolumn{2}{c}{Env.~PM ($\mathrm{km\,s}^{-1}$)}
& \multicolumn{2}{c}{Peculiar PM ($\mathrm{km\,s}^{-1}$)} 
& parallax 
& ref.      
\\
&&&&&
& R.A. & Dec
& R.A. & Dec
& (mas)
&
\\
\midrule
\href{http://simbad.cds.unistra.fr/simbad/sim-id?Ident=\%5BKCS2016\%5D\%20J002153.09\%2B614502.6\%20}{\goneonenine}
& BD\ensuremath{+60^\circ\,39}
& \ra{00}{21}{53}{9}
& \dec{+61}{45}{02.6}
&  	O9V 
& no 
&  -23.81  &  -10.39
&  $ -16.89 \pm -0.14 $  &  $ -7.07 \pm -0.13 $
&  $ 0.41 \pm 0.03 $
& 1 
\\
\href{http://simbad.cds.unistra.fr/simbad/sim-id?Ident=HD+2083}{\hiptwenty}
& BD\ensuremath{+71^\circ\,16}
& \ra{00}{25}{51}{2} 
& \dec{+71}{48}{25.7}
& B1V 
& no 
&  0.19  &  -1.40
&  $ -2.48 \pm -0.16 $  &  $ 6.14 \pm 0.19 $
&  $ 0.98 \pm 0.03 $
& 3
\\
\href{http://simbad.cds.unistra.fr/simbad/sim-id?Ident=%4011672625&Name=TYC%203677-890-1&submit=submit}{\sersix}
& {\tiny 2MASS\;J01112094 +5733282}
& \ra{01}{11}{25}{3}	
& \dec{+57}{34}{0}
& unknown
& no 
&  -12.06  &  -9.83
&  $ -14.59 \pm -0.06 $  &  $ -4.60 \pm -0.04 $
&  $ 0.42 \pm 0.01 $
& 3  \\
\href{http://simbad.cds.unistra.fr/simbad/sim-id?Ident=\%5BKCS2016\%5D\%20J021753.20\%2B611112.8\%20}{\gonethreethree}
& LS\;I\;\ensuremath{+60^\circ\,226}
& \ra{02}{17}{53}{2} 
& \dec{+61}{11}{12.9}
& O8.5-O9IV-V
& no
&  -3.78  &  -8.46
&  $ -4.29 \pm -0.06 $  &  $ 5.94 \pm 0.28 $
&  $ 0.48 \pm 0.01 $
& 1 
\\
\href{http://simbad.cds.unistra.fr/simbad/sim-id?Ident=\%5BKCS2016\%5D\%20J022930.47\%2B612944.1\%20}{\gonethreefour}
& V\ensuremath{^\star}\,KM\,Cas
& \ra{02}{29}{30}{5} 
& \dec{+61}{29}{44.2}
& O9.5V((f))
& no 
&  -1.31  &  -3.66
&  $ -11.92 \pm -0.11 $  &  $ -2.95 \pm -0.05 $
&  $ 0.50 \pm 0.02 $
& 1 
\\
\href{http://simbad.cds.unistra.fr/simbad/sim-id?Ident=\%5BKCS2016\%5D\%20J025410.66\%2B603903.5\%20}{\gonethreeseven}
& BD\ensuremath{+60^\circ\,586}
& \ra{02}{54}{10}{7} 
& \dec{+60}{39}{03.5} 
& O7Vz
& no 
&  1.67  &  -4.37
&  $ -3.96 \pm -0.70 $  &  $ 1.59 \pm 0.23 $
&  $ 0.51 \pm 0.05 $
& 1 
\\
\href{http://simbad.cds.unistra.fr/simbad/sim-id?Ident=\%5BKCS2016\%5D\%20J040553.02\%2B510657.9\%20}{\gonefiveone}
& {\tiny 2MASS\;J04055303 +5106580}
& \ra{04}{05}{53}{0} 
& \dec{+51}{06}{58.0}
& O5V((f))
& no 
&  6.53  &  -18.74
&  $ -11.84 \pm -0.73 $  &  $ -8.82 \pm -0.10 $
&  $ 0.33 \pm 0.02 $
& 1 
\\
\href{http://simbad.cds.unistra.fr/simbad/sim-id?Ident=\%40792954&Name=HD\%20\%2037032}{\hiptwentysix}
& BD\ensuremath{+34^\circ\,1118}
& \ra{05}{37}{09}{0} 
& \dec{+34}{48}{43.1}
& B0.5V
& no 
&  9.93  &  -24.88
&  $ -7.09 \pm -3.94 $  &  $ -2.50 \pm -0.14 $
&  $ 0.42 \pm 0.12 $
& 3 
 \\
\href{http://simbad.cds.unistra.fr/simbad/sim-id?Ident=\%40910971&Name=HD\%20\%2047432}{\hipthirtyone}
& BD\ensuremath{+01^\circ\,1443}
& \ra{06}{38}{38}{2} 
& \dec{+01}{36}{48.7}
& O9.7Ib
& no 
&  -0.48  &  -7.47
&  $ -0.03 \pm -0.02 $  &  $ -4.03 \pm -0.13 $
&  $ 0.59 \pm 0.04 $
& 3 
\\
\href{http://simbad.cds.unistra.fr/simbad/sim-id?Ident=\%40974602&Name=HD\%20\%2057682}{\hdfiftyseven}
& BD\ensuremath{-08^\circ\,1872}
& \ra{07}{22}{02}{1}
& \dec{-08}{58}{45.8}
& O9.2IV
& no 
&  -5.66  &  -0.37
&  $ 56.70 \pm 0.39 $  &  $ 69.93 \pm 0.36 $
&  $ 0.90 \pm 0.07 $
& 3 \\
\href{http://simbad.cds.unistra.fr/simbad/sim-id?Ident=\%5BKCS2016\%5D\%20J163709.54-103401.5\%20}{\zophbs} 
& \zoph
& \ra{16}{37}{09}{5} 
& \dec{-10}{34}{01.5}
& O9.2IVnn
& no 
&  -3.78  &  -9.96
&  $ 10.48 \pm 0.65 $  &  $ 25.79 \pm 0.68 $
&  $ 7.41 \pm 0.66 $
& 1 \\
\href{http://simbad.cds.unistra.fr/simbad/sim-id?Ident=\%402563103&Name=HD\%20165319}{\hdsixteen}
& BD\ensuremath{-14^\circ\,4880}
& \ra{18}{05}{58}{8}	
& \dec{-14}{11}{53.0}
& O9.7Ib
& ? 
&  -9.22  &  -25.99
&  $ -7.47 \pm -0.07 $  &  $ 17.61 \pm 0.35 $
&  $ 0.68 \pm 0.03 $
& 3
\\
\href{http://simbad.cds.unistra.fr/simbad/sim-id?Ident=\%5BKCS2016\%5D\%20J182518.13-130943.0\%20}{\geighteen}
& {\tiny 2MASS\;J18251808 -1309427}
& \ra{18}{25}{18}{1} 
& \dec{-13}{09}{42.7}
& O6V-O5V
& yes
&  -11.76  &  -50.38
& $13.23\pm3.44$ & $21.58\pm0.28$
& $0.24\pm0.02$
& 1 \\
\href{http://simbad.cds.unistra.fr/simbad/sim-id?Ident=\%5BKCS2016\%5D\%20J183821.05-053122.9\%20}{\gtwentysixfifytwo}
& {\tiny 2MASS\;J18382147 -0531233}
& \ra{18}{38}{21}{5}
& \dec{-05}{31}{23.4} 
& unknown
& no
&  n/a  & n/a 
&  n/a  & n/a 
&  $ -0.37 \pm 0.24 $
& 1 
\\
\href{http://simbad.cds.unistra.fr/simbad/sim-id?Ident=\%5BKCS2016\%5D\%20J183909.07-060327.8\%20}{\gtwentysixfourteen}
 & {\tiny 2MASS\;J18390969 -0603279}
& \ra{18}{39}{09}{7} 
& \dec{-06}{03}{27.9}
& unknown
& yes
& n/a  & n/a
& n/a & n/a
& $0.41\pm0.49$
& 1
\\
\href{http://simbad.cds.unistra.fr/simbad/sim-id?Ident=[KCS2016]\%20J184831.03-010209.2}{\gthirtyone}
& {\tiny 2MASS\;J18483100 -0102094}
& \ra{18}{48}{31}{0} 
& \dec{-01}{02}{09.5}
& unknown
& ?
& n/a  &  n/a
& n/a & n/a
& no data
& 1 
\\
\href{http://simbad.cds.unistra.fr/simbad/sim-id?Ident=\%5BKCS2016\%5D\%20J190313.07\%2B053523.0\%20}{\gthirtynine}
& {\tiny 2MASS\;J19031363 +0535230}
& \ra{19}{03}{13}{6}
& \dec{+05}{35}{23.1}
& unknown
& yes 
& n/a  &  n/a
& n/a & n/a
& no data
& 1 \\
\href{http://simbad.cds.unistra.fr/simbad/sim-id?Ident=\%5BKCS2016\%5D\%20J192322.03\%2B144808.4\%20}{\gfortynine}
& {\tiny 2MASS\;J19031363 +0535230}
& \ra{19}{23}{22}{3} 
& \dec{+14}{48}{08.6}
& unknown
& yes 
&  n/a  &  n/a
&  n/a  &  n/a
&  $ -0.13 \pm 0.42 $
& 1 
\\
\href{http://simbad.cds.unistra.fr/simbad/sim-id?Ident=\%5BKCS2016\%5D\%20J202350.71\%2B362426.4\%20}{\gseventyfive}
& {\tiny 2MASS\;J20235070 +3624265}
& \ra{20}{23}{50}{7}
& \dec{+36}{24}{26.6}
& unknown
& ? 
&  n/a  &  n/a
&  n/a  &  n/a
&  $ 0.06 \pm 0.08 $
& 1 
\\
\href{http://simbad.cds.unistra.fr/simbad/sim-id?Ident=\%5BKCS2016\%5D\%20J202717.68\%2B394432.6\%20}{\gseventyeight}
& LS\,II\,\ensuremath{+39^\circ\,53}
& \ra{20}{27}{17}{6} 
& \dec{+39}{44}{32.6}
& O8V(n)((f))
& no 
&  -18.91  &  -38.14
&  $ -7.66 \pm -0.03 $  &  $ -5.00 \pm -0.01 $
&  $ 0.59 \pm 0.01 $
& 1 
\\
\href{http://simbad.cds.unistra.fr/simbad/sim-id?Ident=\%5BKCS2016\%5D\%20J203336.08\%2B435907.4\%20}{\bdfortybs}
& \bdforty
& \ra{20}{33}{36}{1}
& \dec{+43}{59}{07.4}
& O4If
& yes 
&  -19.45  &  -38.01
& $-1.69\pm0.01$ & $43.91\pm0.84$
& $0.58\pm0.01$
& 2
\\
\href{http://simbad.cds.unistra.fr/simbad/sim-id?Ident=\%5BKCS2016\%5D\%20J224640.45\%2B581743.8\%20}{\gonezeroseven}
& BD\ensuremath{+57^\circ\,2606}
& \ra{22}{46}{40}{2} 
& \dec{+58}{17}{44.0}
& B0Ib
& ? 
&  -31.92  &  -22.91
&  $ 0.77 \pm 0.00 $  &  $ -5.05 \pm -0.03 $
&  $ 0.38 \pm 0.01 $
& 1
\\
\href{http://simbad.cds.unistra.fr/simbad/sim-id?Ident=\%5BKCS2016\%5D\%20J225133.77\%2B590030.7\%20}{\gonezeroeight}
& BD\ensuremath{+58^\circ\,2492}
& \ra{22}{51}{33}{8} 
& \dec{+59}{00}{30.8}
& B0
& ?
&  -38.22  &  -26.87
&  $ -60.39 \pm -0.15 $  &  $ -60.73 \pm -0.17 $
&  $ 0.33 \pm 0.02 $
& 1
\\
\href{http://simbad.cds.unistra.fr/simbad/sim-basic?Ident=BD\%2B60+2522}{\bdsixtybs}
& \bdsixty
& \ra{23}{20}{44}{5} 
& \dec{+61}{11}{40.5}
& O6.5(n)fp
& yes 
&  -33.77  &  -20.15
&  $ -4.05 \pm -0.02 $  &  $ 25.65 \pm 1.06 $
&  $ 0.33 \pm 0.02 $
& 4 \\
\bottomrule
\end{tabular}
\tablebib{
(1)~\citet{KobChiSch16}; (2)~\citet{BenRomMar10}; (3)~\citet{PerBenIse15}; (4)~\citet{GreMacHaw19}.
}
\tablefoot{
Columns refer to: (1) bow-shock identifier, (2) star id, (3,4) R.A.~and Dec.~of star, (5) stellar spectral type, (6) radio detection?, (7,8) mean proper motion in the local environment of the star, (9,10) peculiar proper motion of the star with respect to local environment, (11) \textit{Gaia} DR3 parallax, (12) reference for bow shock.
Targets from our survey are named according to the entry in the catalogue of \citet{KobChiSch16} (GXXX.XXXX+XX.XXXX) and the E-BOSS catalogue \citet[][table 8]{PerBenIse15} (EBXX [Star ID]).
The id of the star driving the bow shock is taken first from bright-star catalogues; otherwise the 2MASS identifier is used.
Peculiar proper motions are listed with statistical $1\sigma$ uncertainties only; values with large systematic uncertainties arising from uncertain \textit{Gaia} identification or large parallax uncertainty are listed as 'n/a'.
}
\end{sidewaystable*}

\section{Estimated mass-loss rates and wind velocities for the bow-shock driving stars} \label{sec:app:mdot}

For 17 of the 24 targets we found spectral types in the literature and obtained mass-loss rates, $\dot{M}$, and wind velocities, $v_\infty$, using the method described in Section~\ref{sec:electrondensity}.
These values are listed in Table~\ref{tab:app:mdot}, together with the derived ISM electron density, $n_{e,0}$, from Equation~\ref{eqn:standoff}.
Note that we do not consider a possible radial velocity component to $v_\star$, and so $v_\star$ may be underestimated, leading to an overestimate of $n_{e,0}$.
We also assume the bow shock is oriented perpendicular to our line of sight, which may not be true and introduces some uncertainty in the observational estimate of $R_0$.
Some stars have $v_\star < c_s$ such that a bow shock will not form because stellar motion is not supersonic in the surrounding ISM.
In this case $n_{e,0}$ is not quoted because Equation~\ref{eqn:standoff} does not apply.

\begin{table}[]
    \centering
    \caption{Estimation of mass-loss rates, $\dot{M}$, and wind velocities, $v_\infty$, from spectral types in Table~\ref{tab:targets} of 17 of the 24 target bow shocks and the derived ISM electron density, $n_{e,0}$.}
    \label{tab:app:mdot}
    \begin{tabular}{l|c c c}
    Bow Shock   &  $\dot{M}$ & $v_\infty$ & $n_{e,0}$ \\
      &  ($10^{-6}\,\mathrm{M}_\odot\,\mathrm{yr}^{-1}$) &  (km\,s$^{-1}$) &  (cm$^{-3}$) \\
\midrule
G119.4436-00.9208 	&  0.0617 	& 2384 	&    21.1   \\
EB01: HIP 2036     	&  0.0013 	& 2371 	&     --    \\
EB29: Ser 6       	&  0.1000 	& 2500 	&    14.2   \\
G133.1567+00.0432 	&  0.0997 	& 2361 	&     --    \\
G134.3552+00.8182 	&  0.0435 	& 2267 	&    49.0   \\
G137.4203+01.2792 	&  0.3321 	& 2477 	&     --    \\
G151.0318-00.8271 	&  0.3321 	& 2477 	&    29.1   \\
EB09: HIP 26397    	&  0.0031 	& 2432 	&     --    \\
EB12: HIP 31766    	&  0.0435 	& 2267 	&     --    \\
EB31: HD 57682    	&  0.0663 	& 2314 	&    12.8   \\
$\zeta$ Oph         	&  0.0220 	& 1500 	&     5.3   \\
EB23: HIP 88652    	&  5.3000 	& 2100 	&   120.5   \\
G018.2660-00.2988 	&  1.2885 	& 2639 	&   119.3   \\
G078.2889+00.7829 	&  0.1553 	& 2392 	&     --    \\
EB27: BD+433654     	&  9.0000 	& 2300 	&     9.5   \\
G107.1371-00.6974 	&  0.0910 	& 1890 	&     --    \\
G108.0269-00.3497 	&  0.0910 	& 1890 	&     0.2   \\
NGC\,7635       	&  1.3000 	& 2000 	&    21.9   \\
    \end{tabular}
\end{table}

\section{Sources not detected with the VLA} \label{sec:app:nondetections}

\begin{figure*}
    \centering
    \includegraphics[height=0.24\textwidth]{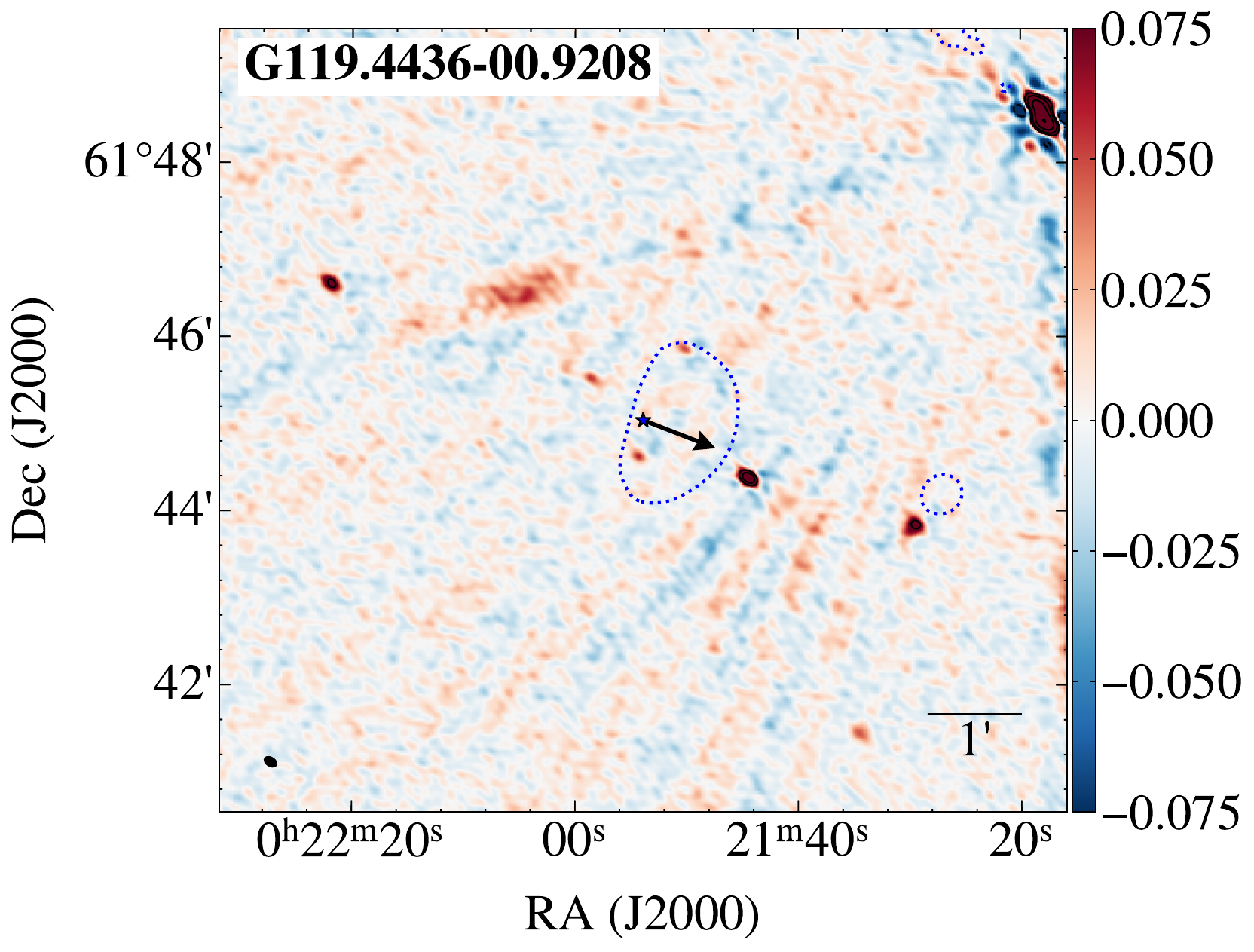}
    \includegraphics[height=0.24\textwidth]{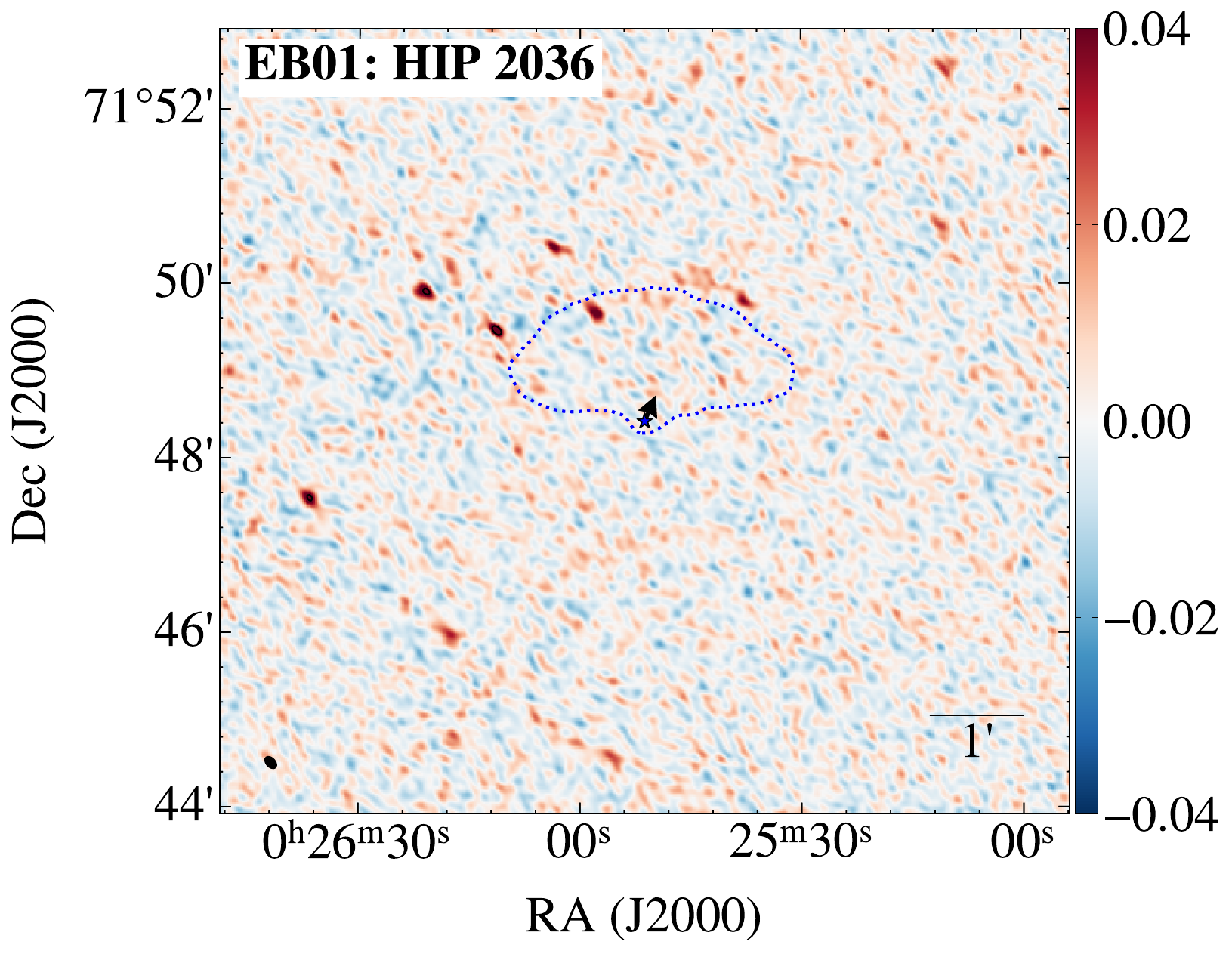}
    \includegraphics[height=0.24\textwidth]{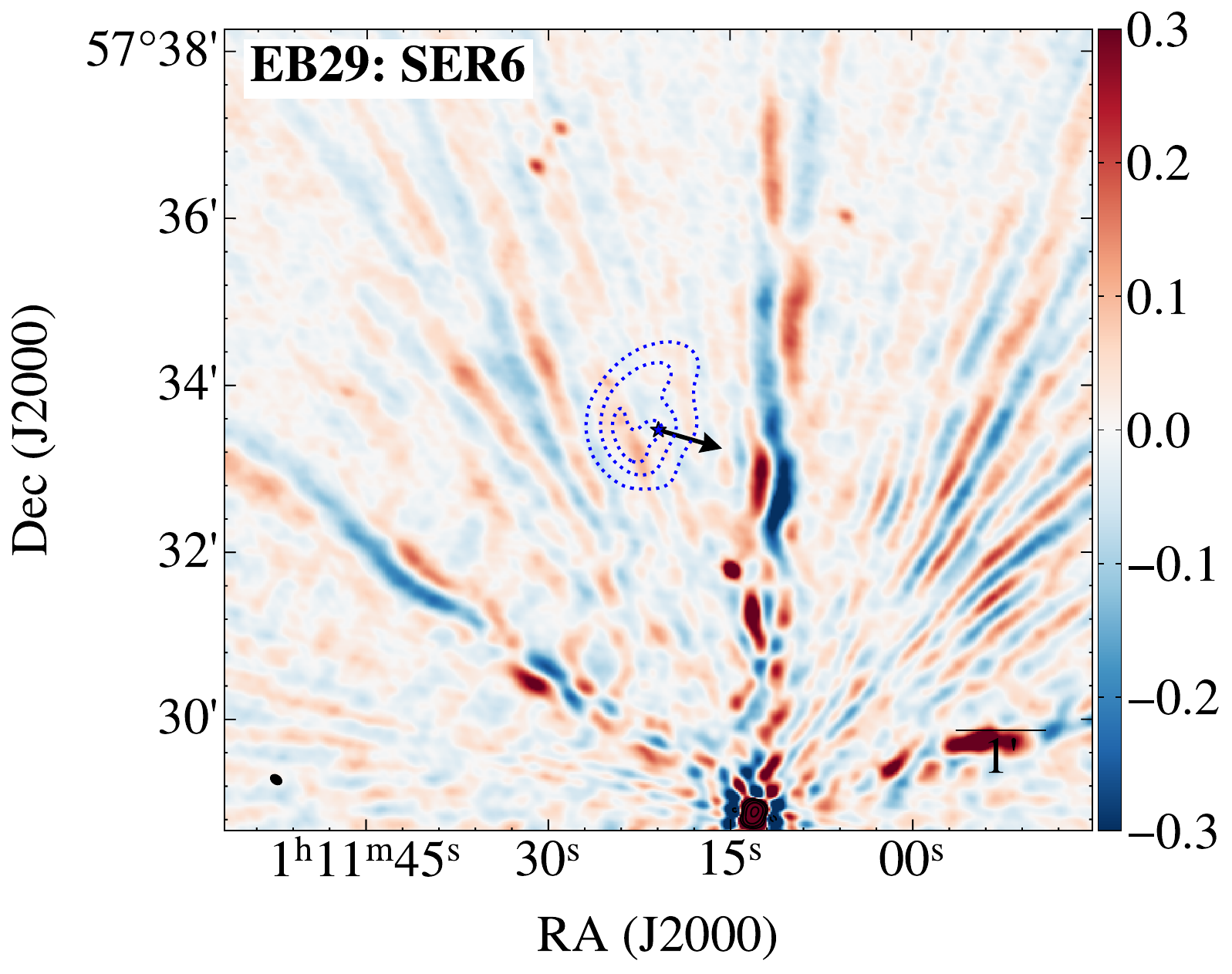}
    \includegraphics[height=0.24\textwidth]{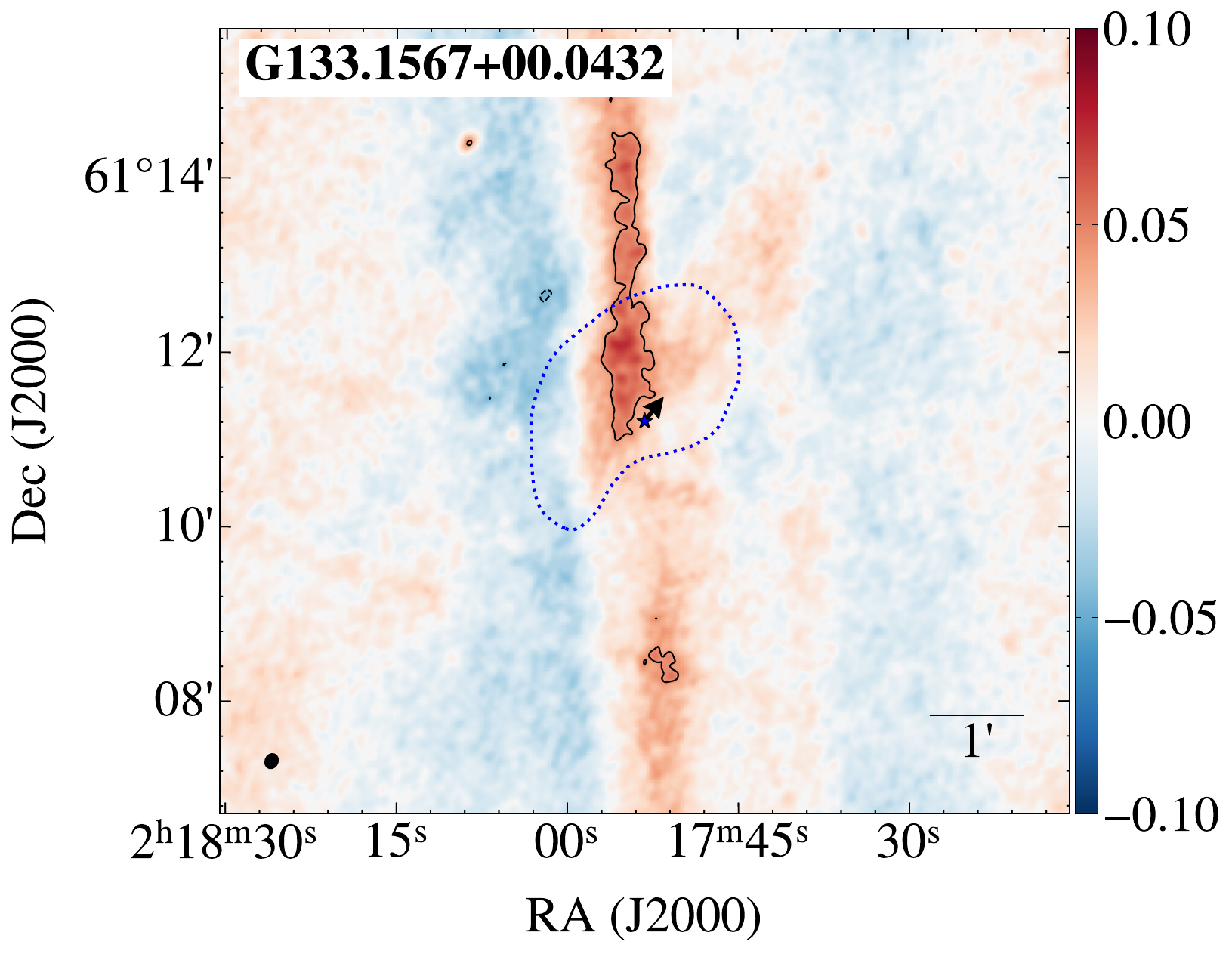}
    \includegraphics[height=0.24\textwidth]{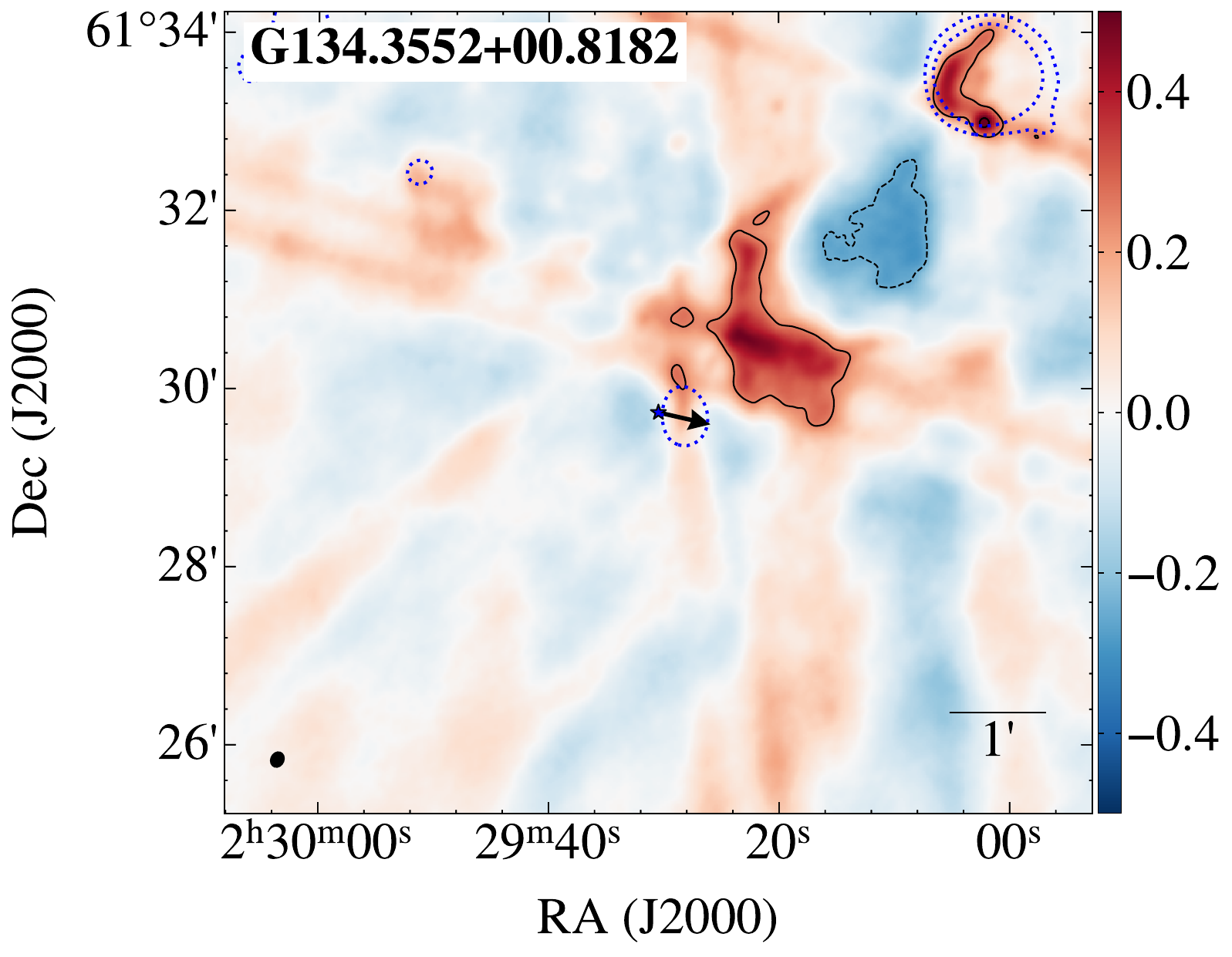}
    \includegraphics[height=0.24\textwidth]{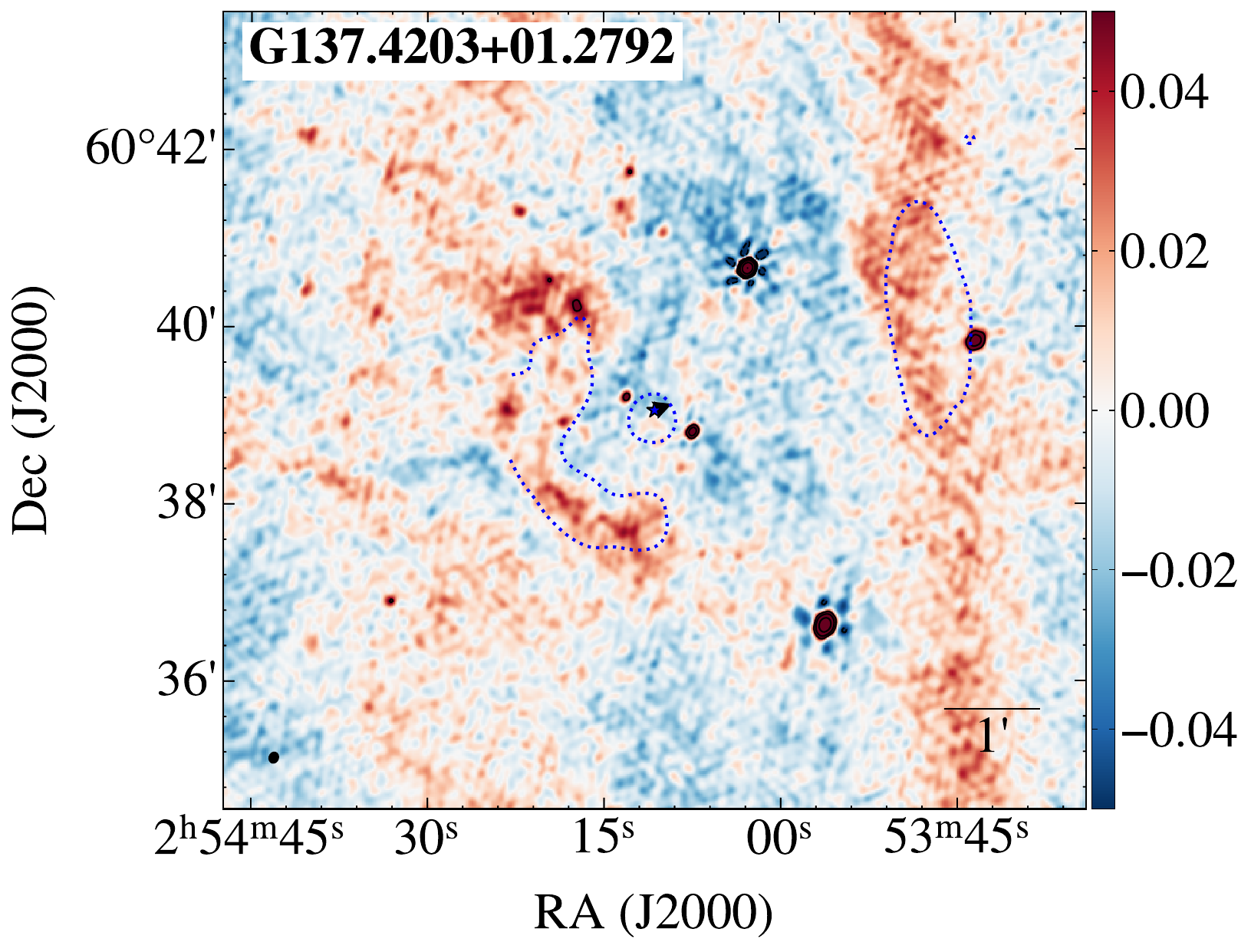}
    \includegraphics[height=0.24\textwidth]{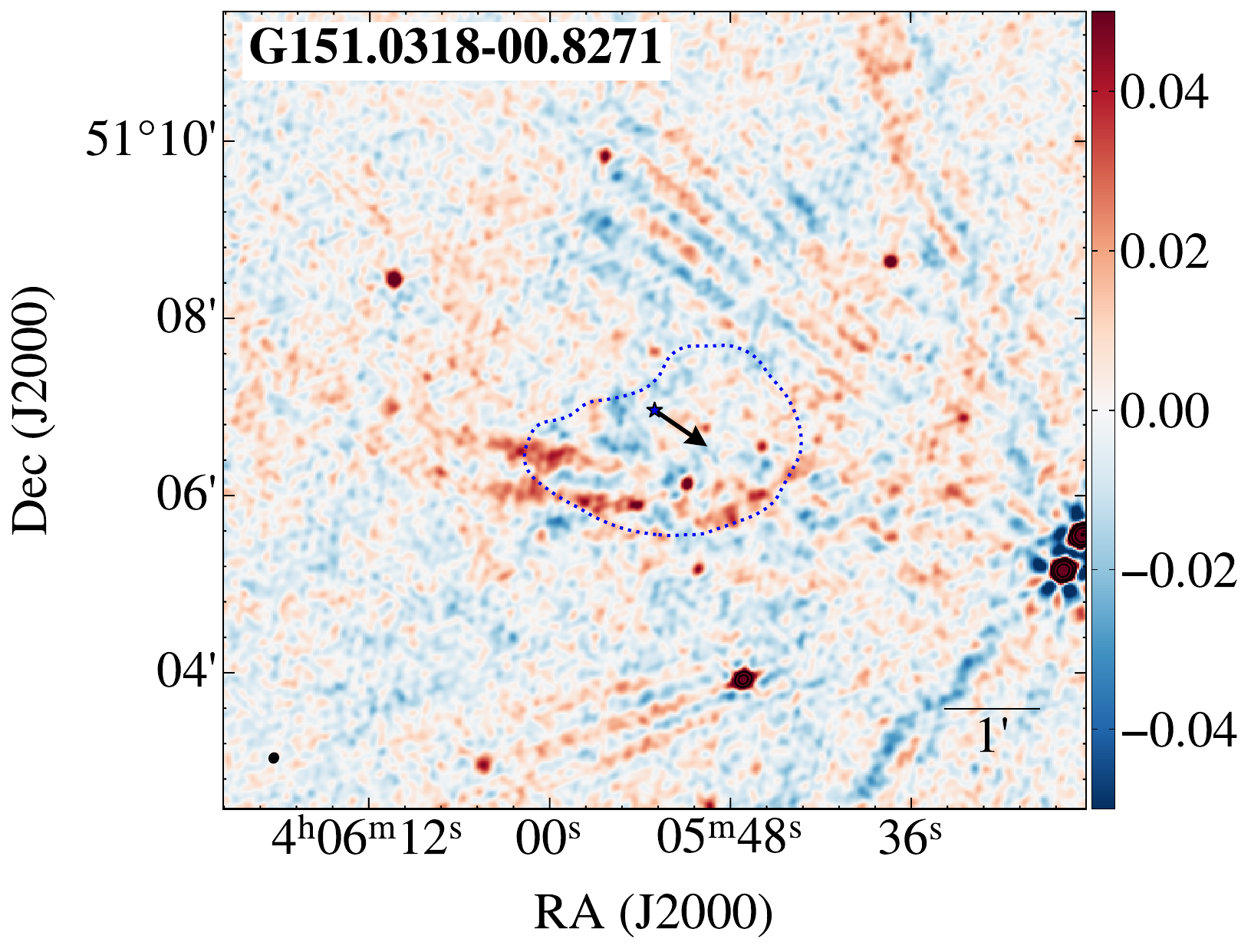}
    \includegraphics[height=0.24\textwidth]{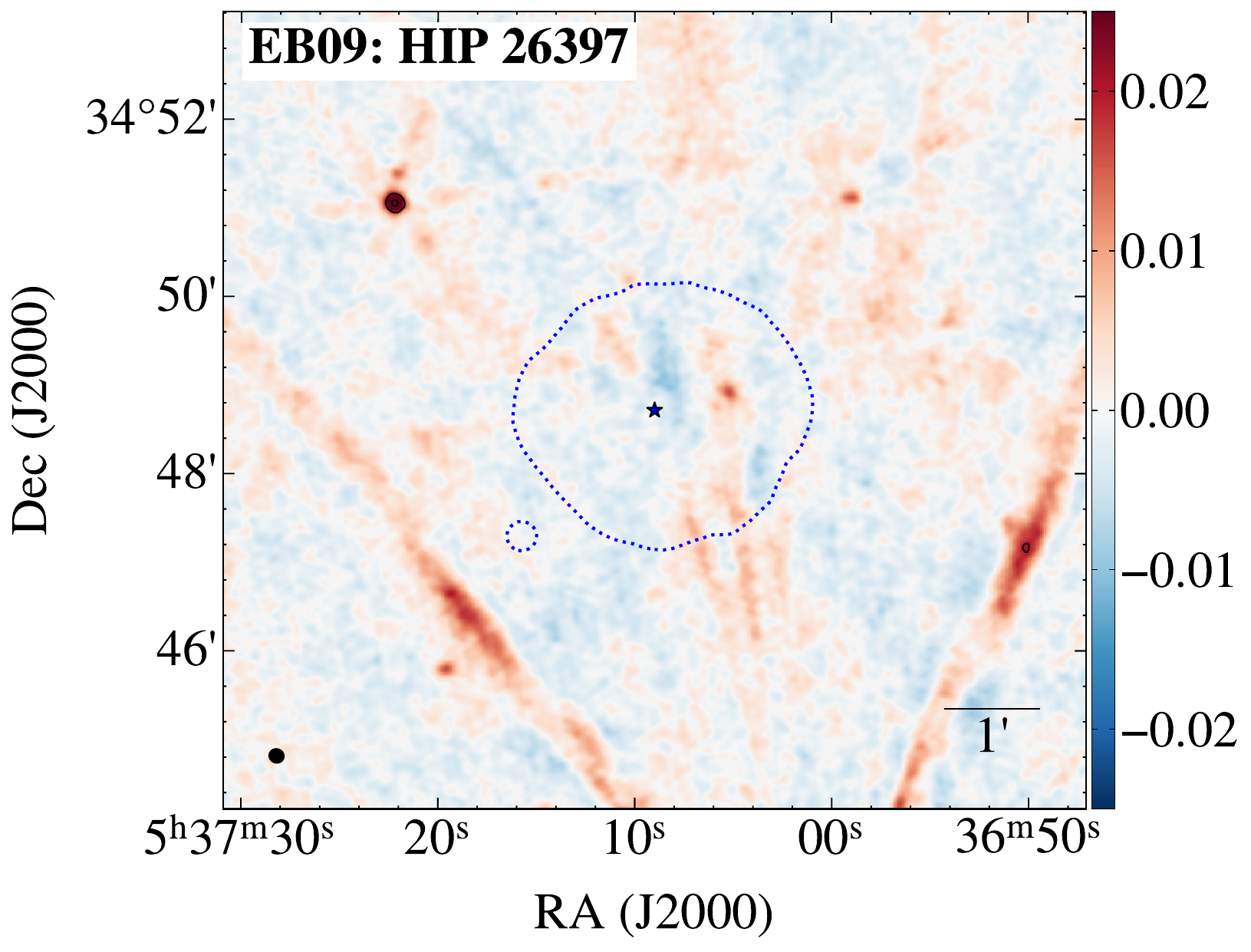}
    \includegraphics[height=0.24\textwidth]{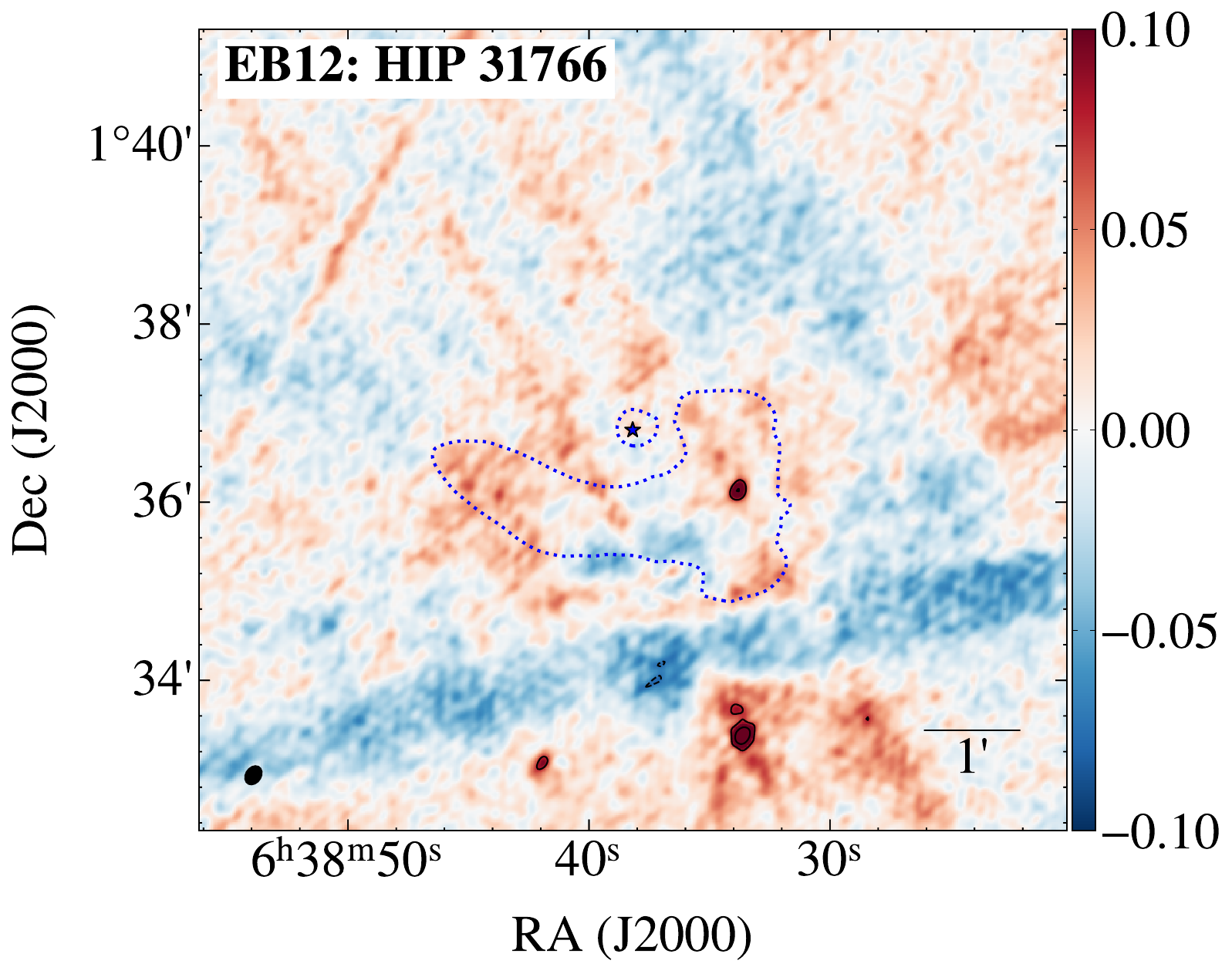}
    \includegraphics[height=0.24\textwidth]{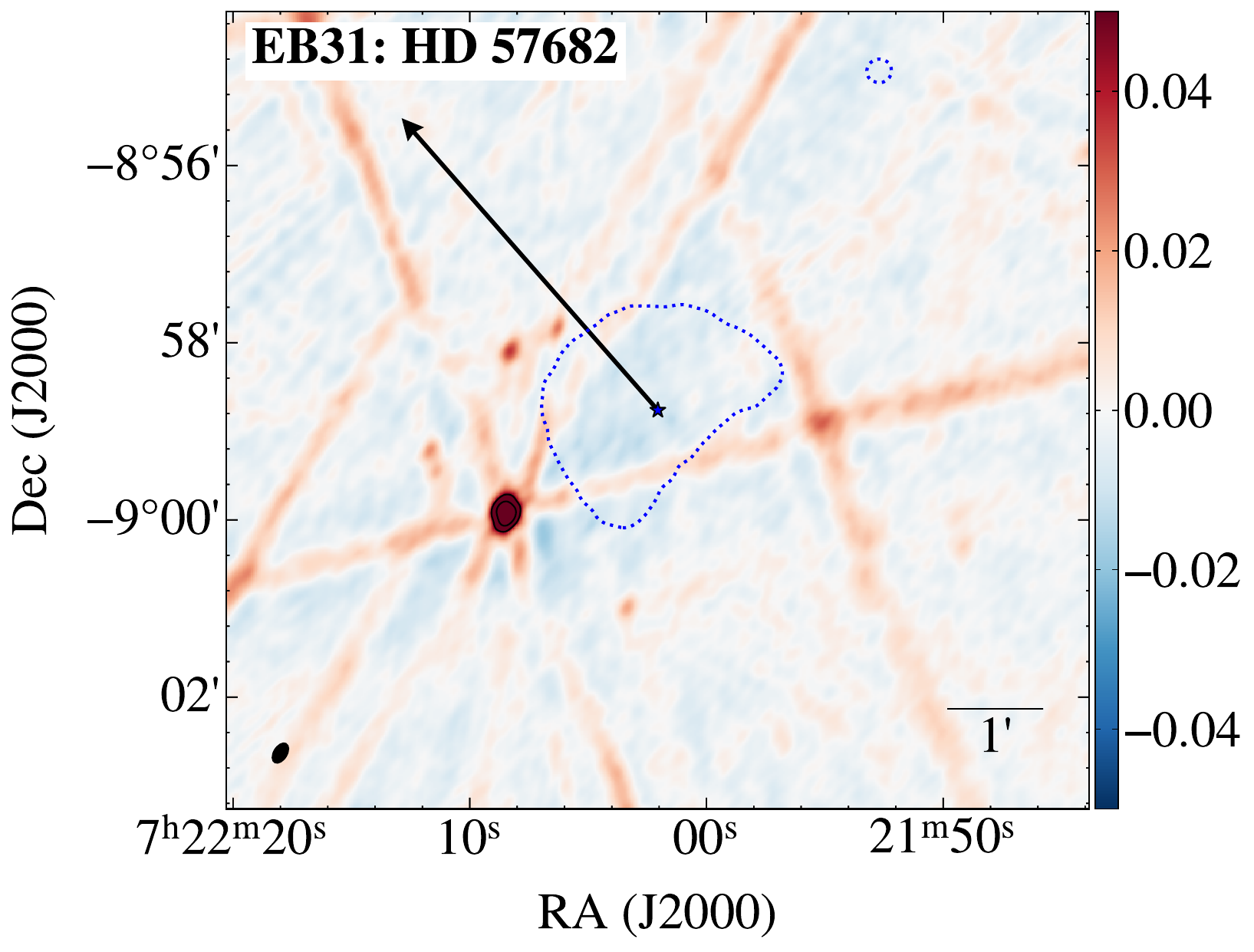}
    \includegraphics[height=0.24\textwidth]{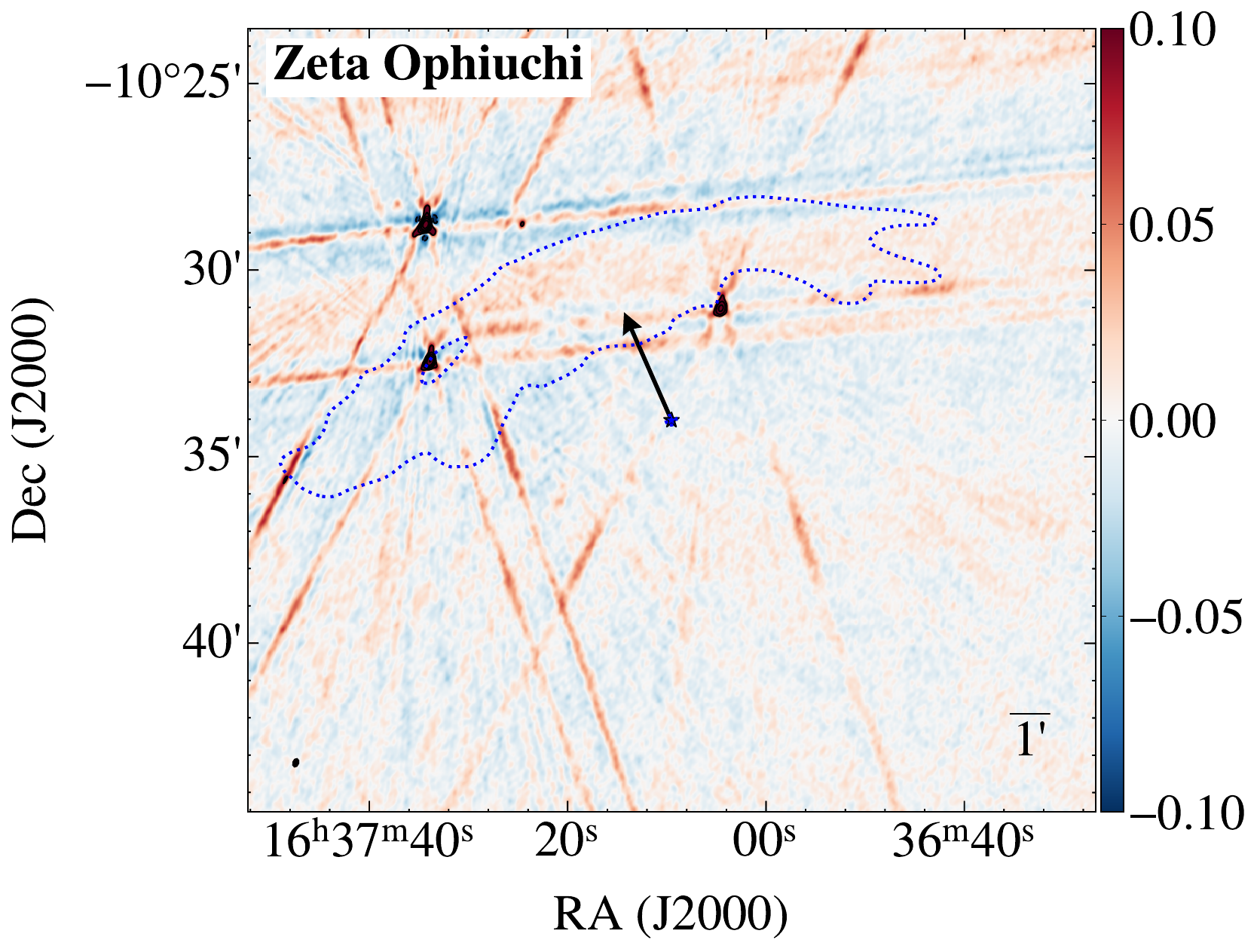}
    \includegraphics[height=0.24\textwidth]{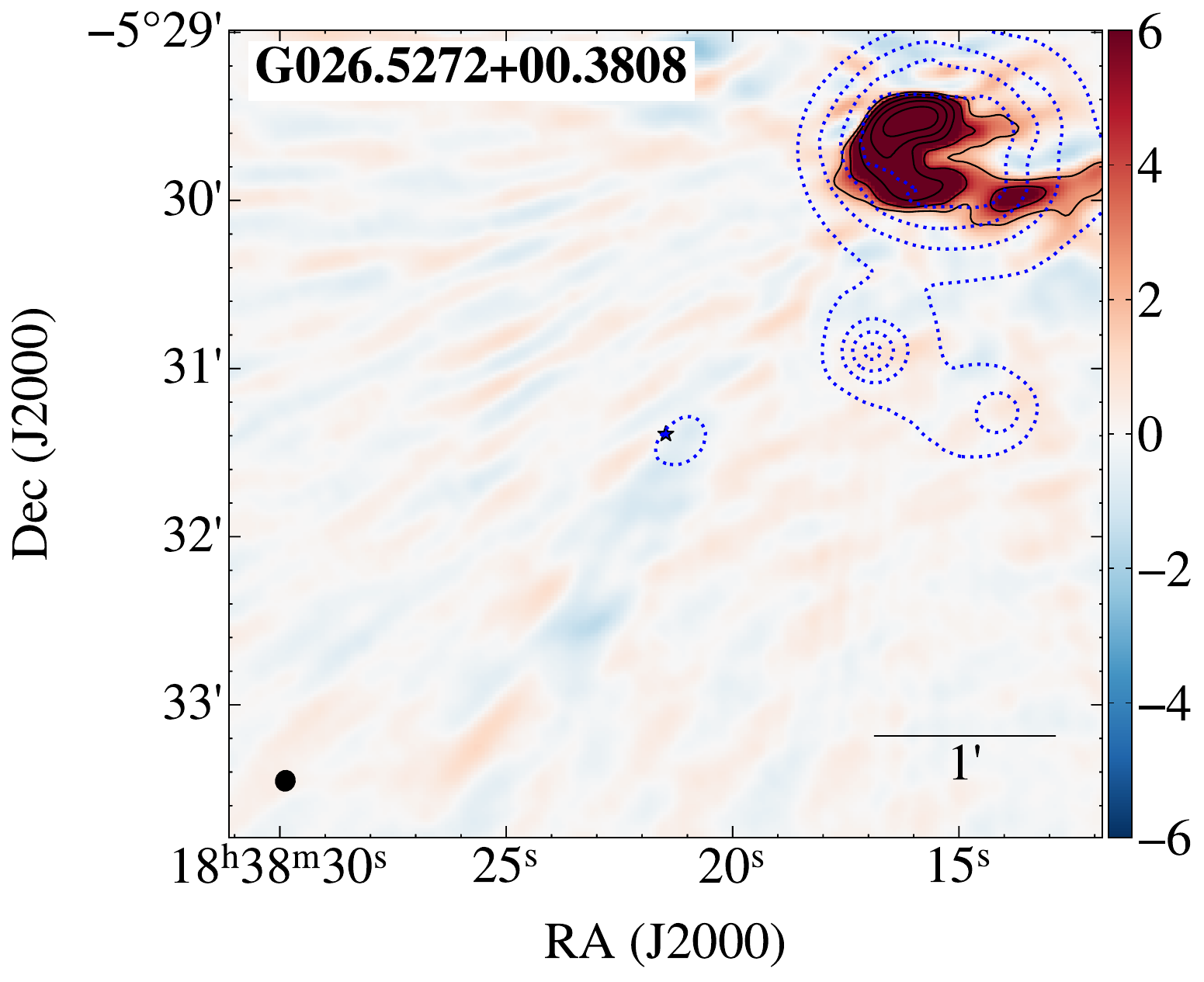}
    \includegraphics[height=0.24\textwidth]{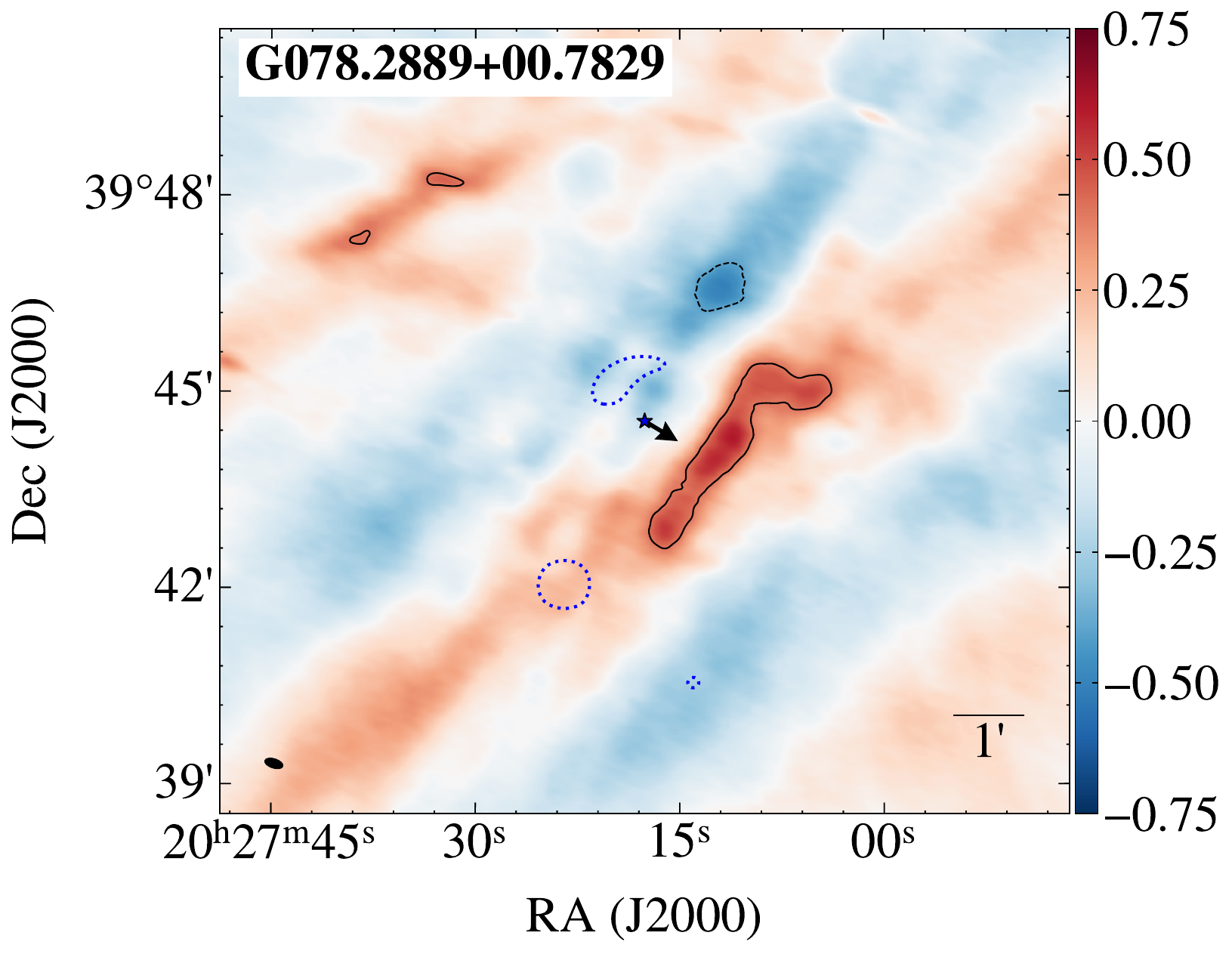}
    \caption{Intensity plots of the non-detected targets listed in Table~\ref{tab:targets}. We use a saturated colour scale extending to negative values to show the noise level in the maps.
    The colour bar values are in milliJansky per beam.
    In the bottom right, a line that corresponds to \ang{;1;} is shown, while in the bottom left, the cross-section of the synthetic beam's FWHM can be seen.
    Contours of the most interesting features from the WISE \SI{22}{\um} counterpart are overlaid with blue lines.
    }
    \label{fig:non-detection-images2}
\end{figure*}

Looking at each of the non-detected targets in turn, and comparing with the literature and the source catalogues, \citet{PerBenIse15} and \citet{KobChiSch16}, we can make the following observations on each source:
\begin{enumerate}
\item
  \goneonenine:
  Entry 364 of the candidate bow shocks in \citet{KobChiSch16} is driven by the O9V star BD\,+60$^\circ$39 and shows an IR arc to the SW of the star.
  Figure~\ref{fig:non-detection-images2} shows that we do detect a moderate peculiar proper motion of the star (indicating a walkaway star) in the direction suggested by the bow shock's orientation, but we do not detect any radio emission coinciding with the IR emission.
  Again this may be due to a lack of sensitivity, or the IR emission could arise from a bow wave that is not accompanied by an increase in the gas density.
\item 
  \hiptwenty:
  Entry 5 of the E-BOSS catalogue of \citet{PerBenBro12}  is a bow shock candidate around the B0.2V star HD\,2083 (aka BD+71$^\circ$\,16 and HIP\,2036) with a faint IR arc to the north of the star.
  The emission is undetected in our VLA observations (Figure~\ref{fig:non-detection-images2}), and nor do we detect a significant peculiar proper motion for the star.
\item 
  \sersix:
  This bow shock candidate from group 7 of the E-BOSS II catalogue \citep{PerBenIse15} is a bright mid-IR arc of emission around the star TYC\;3677-890-1 (aka 2MASS\;J01112094+5733282), which does not have a spectral type reported in Simbad.
  Figure~\ref{fig:non-detection-images2} shows that we detect no diffuse radio emission down to the noise level of the map (which is very noisy compared with most of the other targets because of the bright source to the south).
  There is a small peculiar proper motion for the star (indicating a walkaway star), but in a direction opposite to what would be expected if this source were a bow shock.
\item
  \gonethreethree:
  Entry 367 of the candidate bow shocks in \citet{KobChiSch16} is driven by the late O star LS\,I\;+60\,226.
  We do not detect the bow shock in radio (Figure~\ref{fig:non-detection-images2}), nor do we measure a significant peculiar proper motion for the star ($\approx7$\,km\,s$^{-1}$), although in the IR images it appears to be an excellent candidate.
\item
  \gonethreefour:
  Entry 368 of the candidate bow shocks in \citet{KobChiSch16} is driven by the late O-type variable star V$^\star$\,KM\,Cas, and appears to be within a large H~\textsc{ii} region, with the IR arc in the direction of the closest bright nebulosity.
  It is not detected by the VLA survey (Figure~\ref{fig:non-detection-images2}).
  As such, it could also be an IR arc tracing the edge of a wind bubble around a star moving subsonically with respect to its surroundings \citep{MacHawGva16} and we would not expect to see an excess emission in free-free emission.
  Consistent with this we do not detect large proper motion for the star ($\approx12$\,km\,s$^{-1}$, indicating a walkaway star).
\item
  \gonethreeseven:
  Entry 369 of the candidate bow shocks in \citet{KobChiSch16} is driven by the late O7Vz star BD\,+60$^\circ$586 \citep{SotMaiWal11}, and again appears to be located within a large H~\textsc{ii} region and with the IR arc in the direction of the nearest bright nebular emission to the east.
  There is a vague hint of some emission but it is well within the noise level of the map and the WISE contour in Figure~\ref{fig:non-detection-images2} leads the eye somewhat to see some emission.
  It may be worth following up with deeper observations, although the lack of any significant peculiar proper motion ($<5$\,km\,s$^{-1}$) argues against a bow-shock interpretation of the IR emission.
\item 
  \gonefiveone:
  Entry 371 of the candidate bow shocks in \citet{KobChiSch16} is driven by the O5V((f)) star TYC\,3339-851-1 \citep{RomRom19} (aka 2MASS\;J04055303+5106580) and surrounded by mid-IR nebular emission brightest to the SW.
  Figure~\ref{fig:non-detection-images2} shows that we do not detect any radio emission from the bow shock, but we do measure a small peculiar proper motion for the star appropriate for the bow-shock interpretation  (indicating a walkaway star).
\item
  \hiptwentysix:
  The B0.5V star (aka HD\,37032 and HIP\,26397)  drives bow shock 63 from the E-BOSS catalogue of \citet{PerBenBro12}, which is undetected in our VLA observations (Figure~\ref{fig:non-detection-images2}).
  Interestingly, the IR nebulosity in the cutout image of \citet{PerBenBro12} looks more similar to an Archimedean spiral or a ring nebula than an arc.
  We do detect a small (and uncertain) peculiar proper motion for the central star of $\approx7$\,km\,s$^{-1}$ that is consistent with the orientation of the brightest part of the IR emission.
\item
  \hipthirtyone:
  this star  (aka HD\;47432 and HIP\,31766) is an O9.7I supergiant \citep{SotMaiWal11} and listed as a pulsating variable in Simbad.
  It has what appears to be a double-arc structure in mid-IR emission, pointing to the SW, and is entry 78 in the E-BOSS catalogue of \citet{PerBenBro12}.
  It is undetected in our VLA observations (Figure~\ref{fig:non-detection-images2}), although there is a hemisphere of low-level positive emission that correlates with the position of the IR arc, and so this may be worth following up with deeper observations.
  We also do not detect any significant peculiar proper motion of the star ($\approx4$\,km\,s$^{-1}$), and so it is possible that this is a dusty ejecta nebula and not a bow shock.
\item
  \hdfiftyseven\footnote{the entry EB31 in \citet[][table 8]{PerBenIse15} has a typo and actually refers to HD 57682, not HIP 57862 (P.~Benaglia, private communication).}:
  this star is surrounded by an IR arc detected as entry 103 in the E-BOSS II catalogue of \citet{PerBenIse15}.
  The star (aka HD\;57682) is an O9.7IV sub-giant \citep{SotMaiMor14} and we do not detect any radio emission.
  This could be due to lack of sensitivity or it could be an indication that the IR arc is a bow wave driven by radiation pressure, which leads to an overdensity in dust but not in gas \citep{VanMcC88, HenArt19a}.
  We measure a large peculiar proper motion for the star of $\approx90$\,km\,s$^{-1}$ in the direction consistent with the orientation of the IR arc.
\item
  $\zeta$ Oph:
  the well-known bow shock around the closest O star to Earth, $\zeta$ Oph, has been detected in optical line emission \citep{GulSof79}, IR emission \citep{VanMcC88} and diffuse X-rays from the shocked stellar wind \citep{ToaOskGon16, GreMacKav22}.
  It is, however, a very large bow shock in terms of angular scale and, possibly because of its large angular size, we do not detect it with the VLA (Figs.~\ref{fig:zeta-oph} and \ref{fig:non-detection-images2}).
  As discussed in section~\ref{sec:electrondensity}, the non-detection with Effelsberg gives a slightly weaker constraint statistically, but is significantly more robust because large-scale flux is not lost.
  Our proper motion calculation is in good agreement with other values from the literature \citep[e.g.][]{GvaLanMac12, GreMacKav22}.
\item
  \gtwentysixfifytwo:
  In this field of view we detect bright radio emission from the H~\textsc{ii} region IRAS\,18355-0532 \citep{BroNymMay96}, but the arc-shaped bow shock \citep[candidate 121 in][labelled G026.5272+00.3808]{KobChiSch16} in the direction WSW of the star is not detected (Figure~\ref{fig:non-detection-images2}).
  The star driving the bow shock appears to be 2MASS\;J18382147-0531233. 
  The proper motion that we estimate from \textit{Gaia} DR3 is unreliable because the parallax is negative.
\item
  \gseventyeight:
  this bow shock candidate, entry G078.2889+00.7829 in \citet{KobChiSch16}, surrounds the O8V(n)((f)) star LS\,II\;39\,53 \citep{MaiSotAri16}, which is located within a large H~\textsc{ii} region, and we do not detect any radio emission with the VLA  (Figure~\ref{fig:non-detection-images2}).
  It may be an IR arc from a slowly moving star within a H~\textsc{ii} region that does not produce a bow shock, in which case the emissivity of the arc is enhanced in IR but not in tracers of gas density such as radio free-free emission \citep{MacHawGva16}.
  Consistent with this interpretation, we do not detect a large peculiar stellar motion ($\approx9$\,km\,s$^{-1}$) for the star, and the direction is opposite to what would be expected if the IR arc were a bow shock.
\end{enumerate}

\section{Effelsberg maps of larger-scale emission} \label{sec:app:eff-maps}

\begin{figure*}
    \centering
    \includegraphics[width=0.95\textwidth]{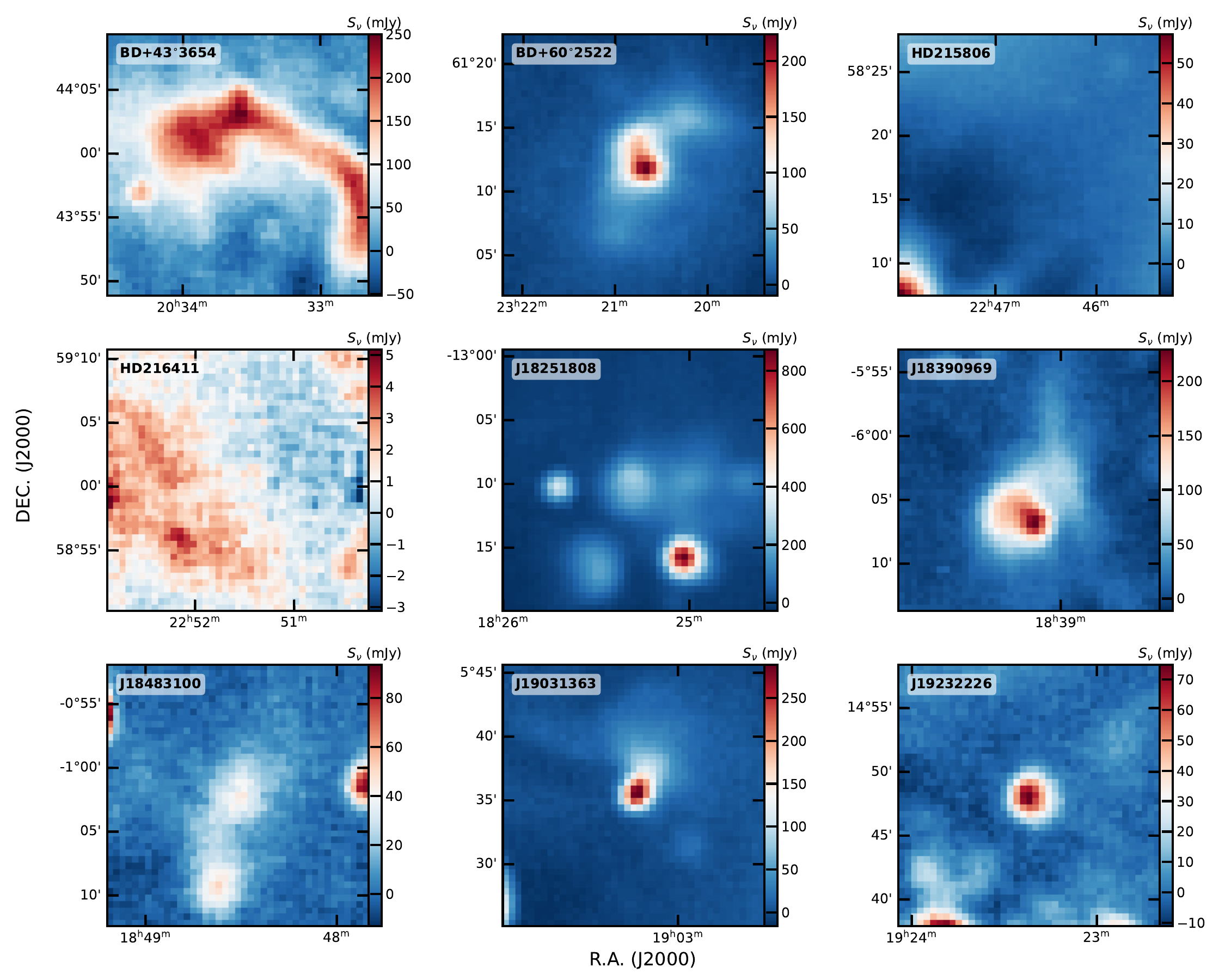}
    \caption{Effelsberg 6.615~GHz continuum images of the nine sources. The source name is labelled in the top left corner of each panel.}
    \label{fig:eff-bow}
\end{figure*}

Figure~\ref{fig:eff-bow} presents the Effelsberg 6.615~GHz radio continuum images for the nine selected sources (all except \zoph{} which is shown in Figure~\ref{fig:zeta-oph}).
While continuum maps were created at seven frequency bands (see Sect.~\ref{sec:effelsberg}), we show only the 6.615~GHz images here as a representative example to illustrate the data quality and morphological features.
The naming convention of the sources is given in Table~\ref{tab:eff-names}

\begin{table}
    \centering
    \caption{List of targets observed with Effelsberg with the associated name in the Effelbserg proposal and in Figure~\ref{fig:eff-bow}.}
    \label{tab:eff-names}
    \begin{tabular}{c|c}
    Target name         &  Name in Effelsberg proposal 86-20 \\
    \hline
    \bdforty            & \bdforty  \\
    \bdsixty            & \bdsixty \\
    \zoph               & \zoph \\
    \gonezeroseven         & HD215806 \\
    \gonezeroeight         & HD216411 \\
    \geighteen          & J18251808 \\
    \gtwentysixfourteen & J18390969 \\
    \gthirtyone         & J18483100 \\
    \gthirtynine        & J19031363 \\
    \gfortynine         & J19232226 \\
    \end{tabular}
\end{table}

\end{appendix}

\end{document}